\documentclass{ieeeaccess}
\usepackage{cite}
\usepackage{caption}
\usepackage{amsmath,amssymb,amsfonts}
\usepackage{algorithmic}
\usepackage{graphicx}
\usepackage{textcomp}
\usepackage[squaren]{SIunits}
\usepackage{xcolor}
\usepackage{listings}
\usepackage{chngcntr}
\usepackage{subfigure}
\usepackage{courier}
\usepackage{hyperref}
\usepackage[hyphenbreaks]{breakurl}

\newcommand{\etal}{\textit{et al}.}

\newcommand{\eg}{\textit{e}.\textit{g}.}
\newcommand{\etc}{\textit{etc}.}
\newcommand{\aka}{\textit{a}.\textit{k}.\textit{a}.}
\newcommand{\FF}{F\&F}
\def\BibTeX{{\rm B\kern-.05em{\sc i\kern-.025em b}\kern-.08em
		T\kern-.1667em\lower.7ex\hbox{E}\kern-.125emX}}

\begin{document}
	\history{Date of publication xxxx 00, 0000, date of current version xxxx 00, 0000.}
	\doi{10.1109/ACCESS.2017.DOI}
	
	\title{Privacy Leakage in Smart Homes and Its Mitigation: IFTTT as a Case Study\\}
	


\author{\uppercase{Rixin Xu}\authorrefmark{1}, \uppercase{Qiang Zeng\authorrefmark{2}, Liehuang Zhu\authorrefmark{3}, Haotian Chi\authorrefmark{4}, and Mohsen Guizani\authorrefmark{5}, \IEEEmembership{Fellow, IEEE}.
}}

	\address[1,3]{School of Computer Science and Technology, Beijing Institute of Technology, Beijing 100081, China, (e-mail: \{xurixin, liehuangz\}@bit.edu.cn)}
	\address[2]{Computer Science and Engineering Department, University of South Carolina, Columbia, SC 29208, USA, (e-mail: zeng1@cse.sc.edu)}
	\address[4]{Department of Computer and Information Sciences, Temple University, Philadelphia, PA 19122, USA,	(e-mail: \{htchi, dux\}@temple.edu)}
	\address[5]{College of Engineering, Qatar University, Doha 2713, Qatar, (e-mail: mguizani@ieee.org)}
	
	\corresp{Corresponding author: Qiang Zeng (e-mail: zeng1@cse.sc.edu).}
	
	\begin{abstract}
		The combination of smart home platforms and automation apps introduces much convenience to smart home users. However, this also brings the potential for privacy leakage. If a smart home platform is permitted to collect all the events of a user day and night, then the platform will learn the behavior patterns of this user before long. In this paper, we investigate how IFTTT, one of the most popular smart home platforms, has the capability of monitoring the daily life of a user in a variety of ways that are hardly noticeable. Moreover, we propose multiple ideas for mitigating privacy leakages, which altogether forms a ``\emph{Filter-and-Fuzz}'' (\FF) process: first, it filters out events unneeded by the IFTTT platform; then, it fuzzes the values and frequencies of the remaining events. We evaluate the \FF~process, and the results show that the proposed solution makes IFTTT unable to recognize any of the user's behavior patterns.
	\end{abstract}
	
	
	\titlepgskip=-15pt
	
	\maketitle
	
	\section{Introduction}
	Smart Home, a typical application of Internet of Things (IoTs), has become increasingly popular in recent years. Smart home devices, such as various sensors and appliances, have been changing the way people interact with their homes. One can monitor remotely monitor \textbf{state information} (\eg, temperature, humidity, occupancy, \etc) or control smart appliances (\eg, turn on/off a lock, configure the routine of a thermostat, \etc) in a smart home. The devices in early stages were heterogeneous, so they could only work in a scattered manner due to limited interoperability.
	Emerging IoT platforms provide a revolution to the smart home. A platform provides a new ecosystem, which typically comprises various smart devices, a local hub, and a backend cloud. Some platforms also provide a programming framework for third-party developers to contribute novel intelligence to smart homes by publishing IoT apps; such platforms are called appified platforms. The users choose IoT apps to control their devices contextually and automatically, known as home automation. Samsung's SmartThings \cite{SmartThi62:online}, Google's Weave/Brillo \cite{WeaveNes79:online}, and Apple's HomeKit \cite{iOSHomeA95:online} are several dominant examples of appified platforms. 
	
	To support more services, devices, and user interfaces, IoT platforms also integrate third-party services by exposing cloud APIs. This allows distinct services, clouds, and applications to manage a smart home collaboratively. 
	
	For instance, SmartThings provides endpoints in its IoT apps (\aka, SmartApps) to allow third-party services/applications (\eg, IFTTT) to gain access to the devices in its system. IFTTT (an initialism for ``\textit{If This, Then That}'') is a free web service to create chains of simple conditional statements which are also called $applets$. ``This'' and ``That'' are the \textit{trigger} and \textit{action} of an applet, respectively. In other words, an IFTTT app (applet) works in the way that ``If a \textit{trigger} is observed, then perform an \textit{action}''. IFTTT can also concatenate different popular Internet services, such as Gmail, Instagram, Facebook, and SmartThings. By integrating SmartThings and IFTTT, users are able to gain more intelligence by installing IoT apps from both SmartThings and IFTTT. 
	
	However, the risk of a privacy breach is also increased by these Trigger-Action IoT platforms. This is because multiple platforms gain access to the users' devices. Despite supporting various services, the device data (\eg, sensor readings, appliance status) are tightly related to user activities and daily routines and the revelation of some sensitive data can cause privacy threats to users.
	
	In this paper, we analyzed the workflows of several typical 3rd-party platforms and found that they share similarities in their potential to monitor a user's daily life excessively in three ways: 1) they can obtain the states of the devices that are not related to any apps; 2) they can get redundant state records, though many records cannot trigger any apps; 3) most third-party apps do not need the accurate values of a numeric sensor measurement, but they continuously receive these values.
	
	We chose to analyze SmartThings (which connects IoT devices and provides services) and some prevalent 3rd-party platforms (which provide services) due to the large user base they have. There are more apps in the SmartThings platform than the competing platforms such as Weavo/Brillo and HomeKit \cite{Fernandes2016}. For IFTTT, there are 11 million users running over 1 billion apps on its server \cite{martin_finnegan_2018}. IFTTT developed a \textbf{Web Service SmartApp} running on SmartThings as an agent, which exposes web endpoints and allows the IFTTT server to access devices in the SmartThings system \cite{smartthingscommunity_2018}.
	
	To prove the redundancy of the event records that are uploaded to 3rd-party platforms, we proposed a mechanism called ``\textbf{Filter\&Fuzz}'' (\FF~for short) to filter the record events. The essential idea of \FF~is that an event does not always have to be uploaded to the 3rd-party apps, and even if it is required, it can be filtered and fuzzed. This significantly reduces the events uploaded to the remote 3rd parties and thus, they can barely recognize a user's behavior pattern. We experimented with \FF~for two agent SmartApps for an identical 3rd-party platform. One is the original agent that monitors and uploads all user events, while the other is customized to only upload events filtered and fuzzed by \FF. The comparative experiment proved that the whole system can still work properly while most event records have been filtered.
	
	However, only filtering the event records is insufficient as the statistical character of the event records can still be calculated to infer users' life patterns. Therefore, other than filtering the records, we proposed a new protocol between the smart home and 3rd-party platforms to hide the true statistical character of a smart home's event records.
	
	Our contributions are summarized as follows:
	
	\begin{enumerate}
		\item We investigate how the integration of several 3rd-party platforms may cause privacy threats to SmartThings users by learning the agent SmartApps of these platforms. We use IFTTT as a representative example to illustrate how these 3rd-party platforms can monitor a user in several ways which are hardly noticeable.
		
		\item We propose a mechanism to prove the redundancy of the event records that uploaded to the 3rd-party platforms. We prove that, the integration of 3rd-party platforms can still work properly if most event records have been filtered.
		
		\item To completely hide the statistical character of the filtered event records, we propose a component for data shared between a major smart home platform and a 3rd-party platform. This component runs on a smart home platform and prevents a Trigger-Action 3rd-party platform from obtaining the pattern of the filtered event records.
	\end{enumerate}
	
	\section{Background and related work}
	
	\subsection{The Architecture of SmartThings Platform}
\begin{figure*}[tbp] \centering
		\includegraphics[width=12cm]{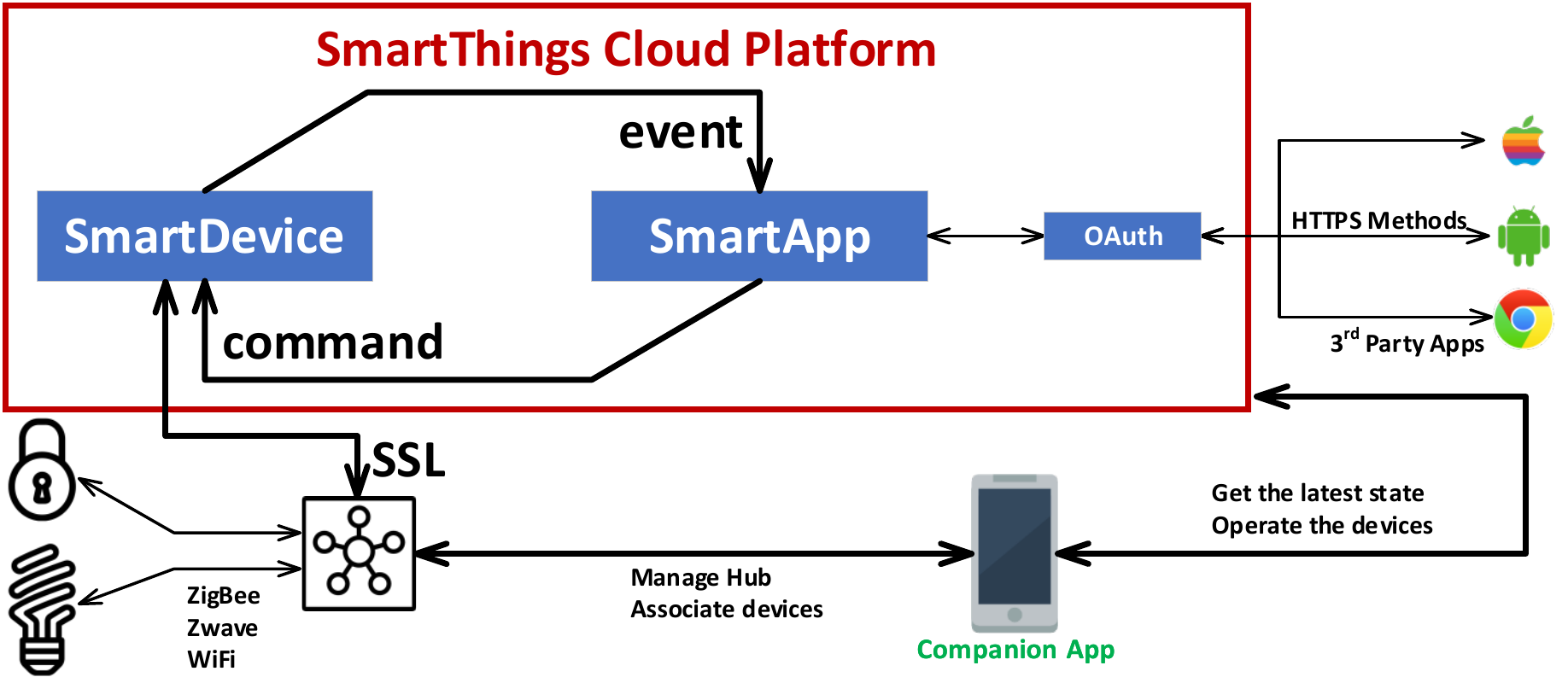}
		\caption{The architecture of SmartThings platform.}
		\label{F_Architecture_SmartThings}
	\end{figure*}

	To use the SmartThings service, a user must buy a SmartThings hub and several end devices. All the end devices are connected via ZigBee, Z-Wave, or Wi-Fi to a hub which maintains an SSL-protected link to the cloud backend. The end devices can be divided into two categories: sensors and actuators. The role of sensors in the SmartThings ecosystem is to gather the state of the house, for example, the \textbf{presence} of the user, the \textbf{illuminance} value of a room, the \textbf{lock} state of a smart lock, the \textbf{power consumption} of an apartment, and so forth. When a sensor detects state changes, the new state will be uploaded to the cloud backend via the hub. These new states are treated as ``events''. Other than sensors, actuators are devices that can ``act''---they perform some specific commands such as ``turning on a switch'' (\texttt{switch.on()}) or ``locking the door'' (\texttt{lock.close()}). Every device in a house is represented by a SmartDevice running on the cloud backend. As the virtual representation of a physical device, SmartDevice translates raw data generated by an end device to events or commands that are suitable for SmartApps. The architecture of the SmartThings platform is shown in Fig.~\ref{F_Architecture_SmartThings}. A user can browse the app market and install SmartApps using the companion app. SmartApps run on the cloud backend, but SmartThings enables the latest hub to run SmartApps. The user can grant a SmartApp a subscription to several end devices, enabling that SmartApp to monitor events generated by sensors and operate actuators by sending commands. For example, a user may install an air conditioner control SmartApp which can be summarized to ``Turn on the air conditioner when the temperature is higher than 30\celsius'' and grant this SmartApp a temperature sensor and an air conditioner. The temperature sensor will upload temperature readings to the cloud, and when this value turns to be above 30\celsius, this SmartApp will send the command to the air conditioner to turn it on. This example also illustrates how SmartThings automatically manipulate devices as the users wish.
	
	Of course, SmartThings have to allow manual operation by the user at any time. Before controlling any devices, the user will refresh for the latest state of his house. This means that the cloud backend must provide timely and accurate state information to the companion app.
	
\subsection{WebService SmartApps and the 3rd-party App}

In order to grab market share and offer more flexibility to developers, SmartThings supports the WebService SmartApps. These SmartApps expose a URL and some defined endpoints, enabling themselves as a tiny web service. A developer can implement a WebService SmartApp and develop a remote 3rd-party app, which runs on a mobile phone or a web server. Granted by a user, the 3rd-party app can obtain an OAuth token released by SmartThings as the credentials to communicate with the corresponding WebService SmartApp via HTTP \texttt{GET}, \texttt{PUT}, \texttt{POST} and \texttt{DELETE} methods. In other words, the 3rd-party app can access the state of, or operate the end devices that are subscribed to the WebService SmartApps with those credentials. The OAuth process between the 3rd-party app and the cloud backend is beyond the scope of this paper. In this scenario, the WebService SmartApp is essentially an \textbf{agent SmartApp} between the 3rd-party app and the end devices. This workflow differs from the regular SmartApps because the developers move the functionality of a regular SmartApp to the 3rd-party app, leaving the WebService SmartApp solely responsible for the communications between the 3rd-party apps and the end devices.

\subsubsection{A 3rd-party app sends messages to the SmartApp}

If a developer wants to enable a 3rd-party app to refresh the state of some devices or send an operation command to an actuator, the \texttt{mappings} declaration in the WebService SmartApp code allows this SmartApp to expose the HTTP endpoints and map the various supported HTTP operations to the associated handlers.


\begin{lstlisting}[caption=Map the HTTP operations to corresponding handlers., frame=shadowbox, label=L_Mapping_code]
mappings {
	path("/switches") {
		action: [
			GET: "listSwitches"
		]
	}

	path("/switches/:command") {
		action: [
			PUT: "updateSwitches"
		]
	}
}
\end{lstlisting}

The snippet in Listing~\ref{L_Mapping_code} shows a typical WebService SmartApp supporting two endpoints. The \texttt{/switches} endpoint will support the \texttt{GET} (line 4) requests. It will call the \texttt{listSwitches()} method and then send back the return value. The \texttt{/switches/:command} endpoint enables the SmartApp to handle the \texttt{PUT} requests (line 10). This differs from from handling the \texttt{GET} requests as it can deliver the \texttt{command} as a parameter to \texttt{updateSwithes()}.	
	
\subsubsection{The WebService SmartApp uploads a new event}
When the WebService SmartApp receives an event, it should send this event to the 3rd-party app immediately. This procedure is also completed by HTTP methods, such as \texttt{httpGet()} and \texttt{httpPostJson()}.

\subsection{SmartThings Working with the Trigger-Action apps}
\label{S_SmartThings_working_with_IFTTT}

Developers, including those from 3rd-parties, Samsung, and the users themselves, have published or deployed numerous apps on various 3rd-party platforms. To authorize a 3rd-party service to his own devices, a user should first grant his 3rd-party account to access ``resources'' of his SmartThings account. Then, SmartThings will automatically deploy a corresponding agent SmartApp to cooperate with the 3rd-party service. Of course, this corresponding agent SmartApp is developed and uploaded by the 3rd-party. After these steps, the 3rd-party service is permitted to communicate with the user's end devices. We selected two popular IFTTT apps which are related to SmartThings as follows.

\begin{enumerate}
	\item \textbf{Turn on Hue Lights when SmartThings detects that you've arrived home} (https://ifttt.com/applets/115637p-turn-on-hue-lights-when-smartthings-detects-that-you-ve-arrived-home, 1.2k users)
	
	\textbf{trigger}: the presence sensor connects to the hub (\texttt{presenceSensor.presence == present})
	
	\textbf{action}: turn on the light (\texttt{switch.on()})
	\item \textbf{Log door openings detected by your SmartThings to Google Drive} (https://ifttt.com/applets/114080p-log-door-openings-detected-by-your-smartthings-to-a-google-drive-spreadsheet, 4.6k users)
	
	\textbf{trigger}: the state of the contact sensor turns to be open (\texttt{contactSensor.contact == open})
	
	\textbf{action}: write the log information to the Google Spreadsheet of the user
\end{enumerate}

As these two apps show, the IFTTT apps can link two different Internet services to work together. If the trigger is a SmartThings end device, then IFTTT will get the latest state of the devices once the state has changed.

\subsection{The Popularity and Necessity of Trigger-Action Apps}

Though there have been numerous SmartApps published on SmartThings, the demands on the automatic functionality varies from user to user and this makes that it quite difficult to find an app which is entirely suitable for a user. Most apps can be described with a Trigger-Condition-Action model: When a new event is triggered, if all the current conditions have been satisfied, then the app will be invoked to execute the action. These sophisticated apps must be created by the developers or teams with sufficient professional knowledge, such as Groovy language and the workflow of SmartThings. However, these prerequisites are extremely difficult for an ordinary smart home user. Therefore, what most users actually need is a platform that is easy to follow and can satisfy as many as various environments as it can.

The Trigger-Action (TA for short) platforms emerged and have been welcomed rapidly; using these platforms does not rely on an amount of technical depth and offers much more flexible choices to the users. With these TA platforms, what a user needs to do is just connect the trigger device and actuators to deploy an app. These apps are quite simple but can cover most cases in daily life. These apps do not need any expert knowledge about the smart home and their functionalities are very intuitive to the user. The TA platforms offer so many conveniences that research on TA 3rd-party platforms is quite neccessary.

\subsection{Related Work}

\subsubsection{IoT security}

Besides the QoS and routing issues \cite{du2005designing, du2006adaptive, du2004qos, mandala2008load}, many efforts also have been put on the security topic of Internet of Things, mobile computing platforms, and smart home and personal privacy for a long time \cite{Islam2012, Lee2014, Lin2016, RisteskaStojkoska2017, cai2018private}. The existing work has proved how vital security is in IoT networks and devices \cite{hei2013pipac}. The current IoT security research mainly analyzes the flaws that come with hardware \cite{Cheng2017}, protocols or key management \cite{Du2007}, and architectures. Sivaraman \etal~analyzed threats and flaws with devices on the market and proposed that SDN technology can be used to block/quarantine and augment the device security of the smart home\cite{Sivaraman2015}. Ali \etal~investigated some potential security attacks in smart homes and evaluated their impacts \cite{Ali2017}. This paper also forecasted that security attacks are expected to be launched in coming years. \cite{Jose2017} proposed a logic based security algorithm to enhance smart home security. The algorithm in this paper is implemented to differentiate normal and suspicious user behavior. Additionally, Blockchain has been applied to enhance the security of the smart home. For example, \cite{Dorri2017} shows an approach to provide decentralized security and privacy and solves overhead issues that are not suitable for resources and power constrained IoT devices. In 2016, Fernandes \etal~first analyzed the security flaws of SmartThings\cite{Fernandes2016}. This paper exploited its framework flaws, including coarse-grained capabilities, insufficient event data protection, flaws in third-party application integration, and unsafe invocation of the groovy dynamic method. Du \etal~proposed the significance of key management of the communications between IoT devices \cite{du2009transactions}. Lee \etal~proposed FACT to handle the coarse-grained capabilities problem in SmartThings \cite{Lee2017} by virtualizing the devices and their functionality. SmartAuth, proposed in \cite{Tian2017}, aimed to check what a SmartApp actually performs and the functionalities it shows to the user. While some malicious operations may be revoked in some unnoticeable scenarios, the ContexIoT was proposed to support the users' fine-grained context identification for some sensitive actions\cite{Jia2017}. To protect the privacy of the user, \cite{Celik2018} proposed how to track sensitive information in SmartThings through static taint analysis of SmartApps.

\subsubsection{The privacy of smart home}

Although manufacturers have deployed various measurements of their platforms, some papers still found flaws leaking users' privacy on various platforms \cite{Geneiatakis2017, Wu2016, liang2018deep, xiao2007survey}. There have been many papers showing how to compromise a user's privacy via the flaws of cloud \cite{Xia2017, Zhou2013, xiao2007internet, du2008security}, protocols, voice interface \cite{zeng2018multiversion}, or even traffic analysis. Yoshigoe \etal~proposed the Synthetic Packet Injection to hide the real traffic between devices and the cloud \cite{yoshigoe2015overcoming}. This paper illustrated the traffic patterns of several SmartThings devices. By the patterns of different devices, an adversary can infer the behavior pattern of a smart home user. Apthorpe \etal~also focused on privacy leakage via network observing \cite{apthorpe2017smart, apthorpe2017spying}. In this paper, they evaluated several strategies to mitigate the risk of privacy leakages when the network traffic is related to the privacy of a user. Then, they proved that traffic shaping can effectively prevent privacy leakages via traffic patterns. But to the best of our knowledge, there are no papers focused on privacy leakage via the 3rd-party apps.

\section{Privacy leakage to IFTTT}

In this section, we will prove and demonstrate how the 3rd-party platform acquires more redundant user data than it actually needs for the functionalities. We use IFTTT as an example to exploit how a 3rd-party platform can monitor the privacy data of users in a way that is easily overlooked.

\subsection{Privacy Leakage by Untrigger-devices}
\label{S_Monitoring_un_trigger_divices}

We treat the devices that cannot trigger any apps as ``\textbf{untrigger-devices}''. There are two types of untrigger-devices: those that have been authorized to the 3rd-party platform but are not related to any apps (we can also call these devices ``idle-devices''), and those related to apps but as the actuators. These actuators only perform the commands of the related apps; they cannot trigger them. The states of these devices are unnecessary to any apps, but the 3rd-party platform will be monitoring all of these devices continuously anyway.

\subsubsection{By the idle-devices}

When a user starts to configure to let a 3rd-party platform access to SmartThings service, they are asked to grant the authorization to the 3rd-party platform. Most users, if not all of them, are willing to authorize all devices to the platform at once, as they are unsure of what kind of apps they will install and which device the new apps might access to in the future. In other words, they are reluctant to go through the tedious authorization process again. However, even if a user triggers device authorization each time when they log in, there still exists a possibility of idleness. Suppose a user is installing an app to be triggered by a switch; when the user deletes this app, the 3rd-party platform will not give up the permission to access to this switch. This is how a new idle-device is created in a way that is easily overlooked.

\subsubsection{By the actuators}

Actuators are another kind of untrigger-device because all they do is passively receive commands from the 3rd-party platform. This means the state change of an actuator is unnecessary for all the 3rd-party apps. But after the authorization, the platform gains the permission to access these actuators. This makes the 3rd-party platform not only able to operate the actuators, but also to monitor the states of these devices.

\subsubsection{Monitoring the untrigger-devices in a practical way}

We reviewed code of the agent SmartApp of IFTTT \cite{smartthingscommunity}, webCoRE \cite{webcorewiki}, and SharpTools \cite{sharptools}, then we got the details of how they monitor the states of untrigger-devices. To exploit this, we used the agent SmartApp code of IFTTT as a representative example.

We start with the HTTP-mapping snippet of the agent SmartApp, as shown in Listing.~\ref{L_IFTTT_mapping_code}. It demonstrates that IFTTT can obtain the states of all devices of the same type at one time. For example, when the agent receives an HTTP message of a \texttt{GET} request with the parameter \texttt{deviceType} as \texttt{humiditySensors}, then the agent will call \texttt{listStates()} and all numerical values of the humidity sensors will be returned to IFTTT. IFTTT currently support 11 kinds of devices: switch, motion sensor, contact sensor, presence sensor, temperature sensor, acceleration sensor, water sensor, light sensor, humidity sensor, alarm, and lock. In daily life, data from these sensors can profile the user's behavior patterns and living environment. But IFTTT does not care if a device can trigger an app or not. Once a device is authorized to IFTTT, it will be monitored whenever IFTTT wants, or even periodically.


\begin{lstlisting}[caption=The HTTP methods mapping snippet of the IFTTT agent SmartApp., frame=shadowbox, label=L_IFTTT_mapping_code]
mappings {
	path("/:deviceType/states") {
		action: [
			GET: "listStates"
		]
}

	path("/:deviceType/:id") {
		action: [
			GET: "show",
			PUT: "update"
		]
	}
}
\end{lstlisting}

\subsection{Leaking by Redundant State Changes}
\label{S_Monitoring_redundant_state}

After checking the monitoring approaches via untrigger-devices, we now focus on ``\textbf{trigger-devices}''. Each IFTTT app has a corresponding trigger statement. Every time the state of a trigger device changes, the agent will upload the new state (namely, an event) to IFTTT. If the trigger of an app is satisfied by the event, it will be invoked and executed.


\begin{lstlisting}[caption=The IFTTT-SmartApp snippet of monitoring and uploading the new events., frame=shadowbox, label=device_handler]
def addSubscription() {
	subscribe(device, attribute, deviceHandler)
}

def deviceHandler(evt) {
	httpPostJson() {...
	}
}
\end{lstlisting}

Listing.~\ref{device_handler} demonstrates the process before the IFTTT agent uploads a new event. Line 2 registers \texttt{deviceHandler()} as the event handler to the \texttt{attribute} of the \texttt{device}. Meanwhile, line 5 shows that the IFTTT agent does not subscribe \texttt{deviceHandler()} to a specific \texttt{attribute} value, but every change of \texttt{attribute}. This can be exemplified by the app ``Turn on Hue Lights when SmartThings detects that you've arrived home''. When the user arrives home, his presence sensor connects to the hub. As a result, the value of \texttt{presenceSensor.presence} will change from \texttt{unpresent} to \texttt{present}. Next, \texttt{deviceHandler()} will be invoked and post \texttt{presenceSensor.presence == present} to IFTTT. At last, the app will be executed. But when the user leaves home in the morning, IFTTT will also receive the event \texttt{presenceSensor.presence == unpresent} because its value changes. However, \texttt{presenceSensor.presence == unpresent} will not trigger any apps. This means that the \texttt{unpresent} is unnecessary for all the apps. In other words, this event is redundant. This also implies that if this redundant event is intercepted before being uploaded to IFTTT, no app will be distracted and all the apps will execute normally as long as the next event can trigger it.  After observing these \texttt{present} events for several days, IFTTT can obtain the approximate arriving time of the user. If a user has no choice but to let IFTTT know the exact arriving time, why should he also let IFTTT know his leaving time?

\subsubsection{Trigger-devices with discrete state values}

A discrete-device is one that has a limited number of possible states. For example, a presence sensor is a typical discrete-device since the value of \texttt{presenceSensor.presence} only contains two possibilities: \texttt{present} and \texttt{unpresent}. Similarly, contact~sensors, locks, and switches are all discrete-devices. In order to protect the privacy of a user, we can specify all the trigger values of these devices for all apps, then the redundant values can be intercepted as they are uploading to a 3rd-party platform.

\subsubsection{Trigger-devices with numeric state values}

Unlike the discrete-devices, the state of a numeric-device covers a range of values. Humidity sensors, illuminance sensors, and temperature sensors are all typical devices that belong to this category. For instance, the state value range of an illuminance sensor may range from 5\lux~(a street lamp) to 100,000\lux~(sunshine). Therefore, the app ``\textit{If the illuminance is exactly 50,000\lux, then ...}'' will not be practically useful. In most cases, triggering this kind of app corresponds to a range, not a specific value. A threshold value divides the possible range into two sub-ranges: the trigger-range and the untrigger-range. Similar to discrete-devices, all state values in the untrigger-range are redundant to the app. But by specifying the trigger-range, we can further protect the user's privacy. Actually, we can hide redundant numeric state values in two aspects. This also can be illustrated by the app ``\textit{If the temperature is above 30\celsius, then turn on the switch.}''

\begin{enumerate}
	\item Suppose the state range of the temperature sensor is [-20, 80]. The threshold value, 30, divides the whole temperature range into two sub-ranges: [-20, 30] and [31, 80].
	\item We can hide changes when the values are below 31\celsius, because these states belong to the untrigger-range and will not trigger this app. This means that the 3rd-party platform will not be able to monitor the temperature fluctuation when it is below 31\celsius.
	\item Even if the latest value is above 30\celsius, we can hide its accurate value. All temperatures above 30\celsius~will trigger the app, regardless of whether it is 35\celsius~or 70\celsius. So, why not transfer the actual temperature state to a random value in the range of [31, 80]?
\end{enumerate}

\subsection{Leaking by the Unnecessary Trigger Value}
\label{F_Monitoring_unnecessary_trigger_value}

Now let us focus on ``that'' of an IFTTT app. The ``that'' part will perform operations like ``call my phone'' or ``play a song from Google music.'' It can also operate a SmartThings actuator. When an app operating an actuator is triggered, it will send an HTTP \texttt{PUT} message to the agent with the corresponding command parameter. Then, the agent operates the actuator device. Additionally, the actuator can be operated automatically or manually by the user. No matter how the actuator is operated, it will upload its new state to the SmartThings backend if the operation changes its state. But sometimes the SmartApp command or manual operation does not change the state of a device, and the device will simply discard this command. Let us use the app, ``\textit{If the temperature is above 30\celsius, then turn on the switch}'' as an example again. The switch can be automatically or manually turned on by the SmartApp or the user. When a switch has been turned on, it receives a \texttt{switch.on()} command and the switch does not need to repeat another \texttt{switch.on()} operation, it will discard it. The \texttt{switch.on()} command in this case is unnecessary as it changes nothing. This implies that the app does not have to be invoked. Therefore, we can conclude that if the current state of the actuator for an app equals the \textbf{consequential state value} of the command, then the app does not have to be invoked and the trigger event can be intercepted before it will be uploaded to IFTTT. But now even if all the events have been filtered or randomized as we have described at Sec.~\ref{S_Monitoring_un_trigger_divices} and Sec.~\ref{S_Monitoring_redundant_state}, there will be still too many redundant events uploaded to IFTTT.

We also investigated the agent SmartApp code of webCoRE and SharpTools; it has been proved that there also exists potential privacy leakage like IFTTT in these two platforms.

\section{\FF: the filtering component}
\label{S_Specific-fuzzification}

In this section, we will firstly introduce how \FF~filters redundant events. As for how to eliminate the statistical character of the events, it will be illustrated in Sec. \ref{fuzz}.

\subsection{The Key Information Items of Every App}

We could extract key items that reflect how an IFTTT app is triggered and what it will do. These items include the \texttt{trigger device}, \texttt{trigger state} or \texttt{state range} for a numeric-device, the \texttt{actuator} (if any), and the \texttt{consequential state value of the actuator} (if any, CSV for short). There is a one-to-one mapping between a 3rd-party app and the combination of all key information. For example, all the key information of ``\textit{If the temperature is above 30\celsius, then turn on the switch}'' can be summarized as \{\texttt{trigger device}: temperature sensor, \texttt{trigger state (range)}: [31, 100], \texttt{actuator}: switch, \texttt{the CSV of the actuator}: on\}. The first step of \FF~is to extract the key information from the apps that are in use. If there are $D$ devices subscribed to a user who has installed $A$ apps, then we will obtain the corresponding $A$ records, and $A \leq D$.

\subsection{Extract the Key Information via a Chrome Extension}

\begin{figure}[tbp] \centering
	\includegraphics[width=0.95\columnwidth]{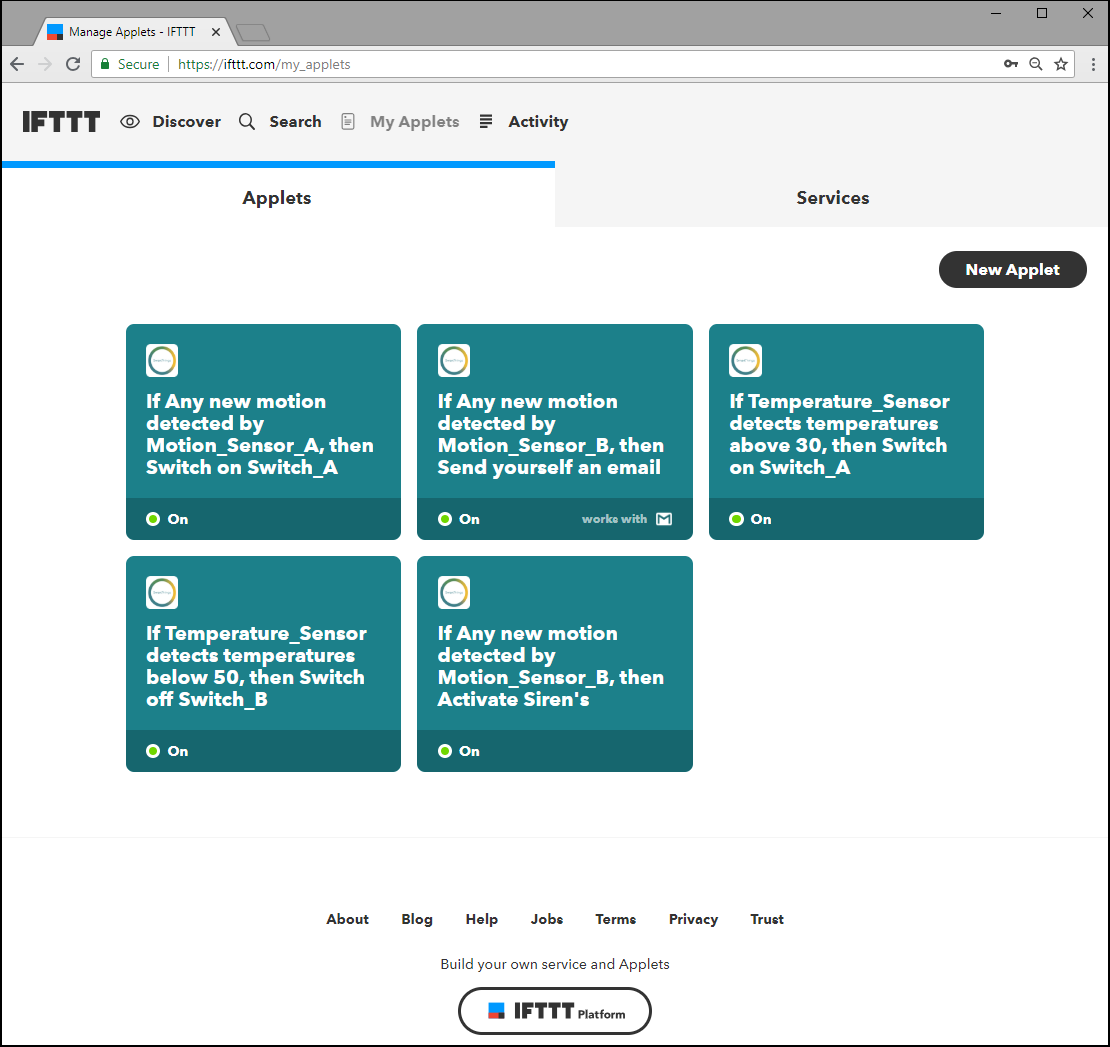}
	\caption{The IFTTT website displays all the apps of a user.}
	\label{F_IFTTT_display_applets}
\end{figure}

A user can manage his 3rd-party service configuration via a mobile app or a desktop web browser. Take IFTTT as an example; all apps can be displayed in the page ``https://ifttt.com/my\_applets'' (Fig.~\ref{F_IFTTT_display_applets}). Each app in the browser is labeled by a one-sentence description, from which the user can catch the functionality of an app at a glance. These label sentences are embedded in the HTML source code of the ``/my\_applets'' page. We can find these descriptive sentences in the \texttt{span} tags with the class value ``\texttt{title}''. The corresponding HTML snippet is shown in Listing~\ref{L_HTML_code}.

To IFTTT, we developed a Chrome extension to extract these sentences. This Chrome app can transfer all the sentences to a new form where every descriptive sentence is split into key information items. To the other 3rd-party services, they also provide a web-based configuration interface. Therefore, all the key information items can also be extracted in this way.


\begin{lstlisting}[caption=The descriptive sentences of all the IFTTT apps can be found in the HTML code., frame=shadowbox, label=L_HTML_code]
<!DOCTYPE html>
<html>
	<span class="title">
		If Any new motion detected by Motion_Sensor_A, then Switch on Switch_A
	</span>
</html>
\end{lstlisting}

\subsection{Merge the Key Information Collection}

This step is transforming the $A$ records to a new form where all records will be merged by identical trigger-devices. We denote the list of trigger-devices as $T$, and $|T|\leq A$ because sometimes a trigger-device can be subscribed by several apps. Suppose that the lists of discrete-devices and numeric-devices are $T_d$ and $T_n$ respectively, then $|T| = |T_d| + |T_n|$.

\subsubsection{Merge the A records of discrete-devices}

A specific state value of a discrete-device may trigger zero, one or multiple apps. To a discrete-device, we assume that this device corresponds to $a$ actuators. All potential state values of this discrete-device form a set $S$ with $p$ elements, but only $t$ of these elements can trigger an app, and $t \leq p$. After checking these $a$ apps triggered by this device, a table can be generated to show the mapping between every value of $S$ in a list, which consists of the consequential state of every actuator, shown in Table.~\ref{T_Map_state_values_to_actuators}. Every row of this table contains at least one element. For each trigger and discrete-device, a similar table can be generated.

\begin{table}[tbp]
	\caption{Map the state values of a discrete-device to the consequential states of the actuators that it can trigger.}
	\begin{center}
		\begin{tabular}{|c|c|c|c|}
			\hline
			& \boldmath{$Actuator_1$}                         & \textbf{...} & \boldmath{$Actuator_a$}                         \\ \hline
			$S_1$                               & \begin{tabular}[c]{@{}c@{}}CSV of $Actuator_1$\\ (if exists)\end{tabular} & ...                                    & \begin{tabular}[c]{@{}c@{}}CSV of $Actuator_a$\\ (if exists)\end{tabular} \\ \hline
			...                                 & ...                                                                       & ...                                    & ...                                                                       \\ \hline
			$S_t$                               & \begin{tabular}[c]{@{}c@{}}CSV of $Actuator_1$\\ (if exists)\end{tabular} & ...                                    & \begin{tabular}[c]{@{}c@{}}CSV of $Actuator_a$\\ (if exists)\end{tabular} \\ \hline
			...                                 & \texttimes                                                 & ...                                    & \texttimes                                                 \\ \hline
			$S_p$                               & \texttimes                                                 & ...                                    & \texttimes                                                 \\ \hline
		\end{tabular}
		\label{T_Map_state_values_to_actuators}
	\end{center}
\end{table}

\subsubsection{Merge the records of numeric-devices}

A numeric-device triggers apps with a range rather than a specific value. Like discrete-devices, the current value of a numeric-device may trigger zero, one, or multiple actuators. For example: ``If Illuminance sensor detects brightness above 500\lux, then switch off $Switch_A$'' and ``If Illuminance sensor detects brightness above 600\lux, then switch off $Switch_B$.''

Like discrete-devices, the state of a numeric-device may trigger the operation of one or multiple devices at the same time (\eg, the brightness is 650\lux), or trigger no app (\eg, the brightness is 400\lux). The corresponding app may be triggered or not depending on whether the current state value is above or below a threshold value. After checking all the apps that are triggered by a specific numeric-device, \FF~will take threshold values and split the measuring range of this device into several sub-ranges. If there are $t$ threshold values, \FF~will arrange these $t$ values in ascending order to form a list. Each element in this list is denoted as $L_i$ ($1\leqslant i\leqslant t$), and $L_1\leqslant...\leqslant L_t$. Then the measuring range covering $min$ to $max$ will be divided into $t+1$ sub-ranges: $[min, L_1], [L_1 + 1, L_2] ... [L_t + 1, max]$. We also can generate a table for a trigger and numeric-device, as shown in Table.~\ref{T_Map_numeric_values_to_actuators}. For each trigger and numeric-device, a similar table can be generated.

\begin{table}[tbp]
	\caption{Map the numeric range of a numeric-device to the consequential states of the actuators that it can trigger.}
	\begin{center}
		\begin{tabular}{|c|c|c|c|}
			\hline
			\textbf{}           & \boldmath{$Actuator_1$}                                                      & \textbf{...} & \boldmath{$Actuator_a$}                                                     \\ \hline
			$[min, L_1]$     & \begin{tabular}[c]{@{}c@{}}CSV of Actuator\_1\\  (if exists)\end{tabular} & ...          & \begin{tabular}[c]{@{}c@{}}CSV of Actuator\_a\\ (if exists)\end{tabular} \\ \hline
			...                 & ...                                                                       & ...          & ...                                                                      \\ \hline
			$[L_t + 1, max]$ & \begin{tabular}[c]{@{}c@{}}CSV of Actuator\_1\\ (if exists)\end{tabular}  & ...          & \begin{tabular}[c]{@{}c@{}}CSV of Actuator\_a\\ (if exists)\end{tabular} \\ \hline
		\end{tabular}
		\label{T_Map_numeric_values_to_actuators}
	\end{center}
\end{table}

\subsection{Filter and Randomize Procedure}

After extracting key information from apps and merging items, the \FF~can begin working. The procedure of \FF~consists of intercepting the event from an untrigger-device, intercepting events that cannot trigger any apps, intercepting events can unnecessarily trigger an app, and finally, randomizing values within a numeric range.

\begin{enumerate}
	\item When a new event comes to the agent SmartApp of \FF, the agent first checks if the event is from a trigger-device or not. If the corresponding device is not in $T$, then all events from this device should not be uploaded.
	\item \FF~has constructed the trigger-state lists for each trigger-device. After the first step, although this new event comes from a device that belongs to $T$, if it does not belong to the trigger-state lists of this device, it will also not be uploaded.
	\item After the above two steps, \FF~will check the current state value of the corresponding actuator (if there is a corresponding SmartThings actuator device). If the current state value equals the consequential state value of the app, it means that this event will change nothing if it is uploaded, and this event will be intercepted too.
	\item Finally, if the event is to be uploaded, \FF~will check if the event is numeric or not. If it is a numeric event, \FF~will firstly check the sub-range of this value (Table.~\ref{T_Map_numeric_values_to_actuators}) and generate a random value within the sub-range.
\end{enumerate}

An event will not be uploaded until it has been checked by these steps.

\section{Redundancy of the original event records}

In this section, we prove the redundancy of the event records that are uploaded to the 3rd-party platforms. With the filtering feature of \FF, the amount of the events that are uploaded to a 3rd-party platforms will be decreased rapidly.

\subsection{CASAS Datasets Overview}

We analyzed four \textit{CASAS} \cite{casasdatasets} datasets: \textit{hh104}, \textit{hh105}, \textit{hh110}, and \textit{hh111}. These datasets are the monitoring records of different users living in a single apartment over two months, except \textit{hh110}, which is for about one month. The event records in each single-apartment include events generated by the \textit{motion sensors}, \textit{contact sensors}, \textit{temperature sensors}, \textit{switches}, and \textit{remaining battery} of some devices.

\subsection{Set the Applets for Every Dataset}

We divided all the devices of every dataset into two groups: trigger-devices and idle-devices. In the trigger-device group, \textit{motion sensors} and \textit{contact sensors} are discrete-devices, and \textit{temperature sensors} are numeric-devices. In the idle-device group, there is an actuator sub-group. We used \textit{switches} as actuators because they can be operated both automatically or manually as mentioned above.

We set one app for every \textit{switch}, and randomly selected devices from the trigger-device group for every app. Then, all remaining devices are idle-devices.

\begin{enumerate}
	\item For discrete-devices, we set the apps as: if the value of motion sensor is ``active,'' or if the value of contact sensor is ``open'', then the corresponding switch is turned on.
	
	\item We calculated the average value of every \textit{temperature sensor} and set the rules triggered by these \textit{temperature sensors} as: if the temperature is above its average value, then the corresponding switch is turned on.
\end{enumerate}

\begin{table}[]
	\centering
	\caption{Comparing the amount of event records of original and filtered datasets (all database starts at 2011-06-15)}
	\label{T_Comaring_the_amount_two_IFTTTs}
	\begin{tabular}{|c|c|c|c|c|c|c|}
		\hline
		\textbf{}        & \textbf{hh104} & \textbf{hh105} & \textbf{hh110} & \textbf{hh111} \\ \hline
		Amount of devices & 26             & 19             & 15             & 22             \\ \hline
		Trigger-devices   & 6              & 6              & 7              & 8              \\ \hline
		Original datasets             & 125900         & 91055          & 41802          & 100896         \\ \hline
		Datasets filtered by \FF    & 2831           & 1317           & 972            & 1599           \\ \hline
	\end{tabular}
\end{table}


For each \textit{CASAS} dataset, we compare the amount of event records before and after filtering by \FF.

\subsection{Comparing the Records Amount}

Each event record of a user contains behavior pattern information for that user. The more event records IFTTT gets, the more likely IFTTT is able to discover and recognize the behavior pattern of a user. So, this leads to one feature of \FF: filtering as many events are uploaded to IFTTT as possible. We experimented with the datasets and compared the results before and after filtering \FF. This is can be shown in Table.~\ref{T_Comaring_the_amount_two_IFTTTs}. As we can see, for the same dataset, the \FF~ significantly reduces the amount of the records significantly compared to the original dataset. The amount of event records uploaded to the 3rd-party platform is no more than 2.2\% (\textit{hh104}) of the original dataset.

\begin{figure}[tbp] \centering
	\subfigure[] {
		\includegraphics[width=0.45\columnwidth]{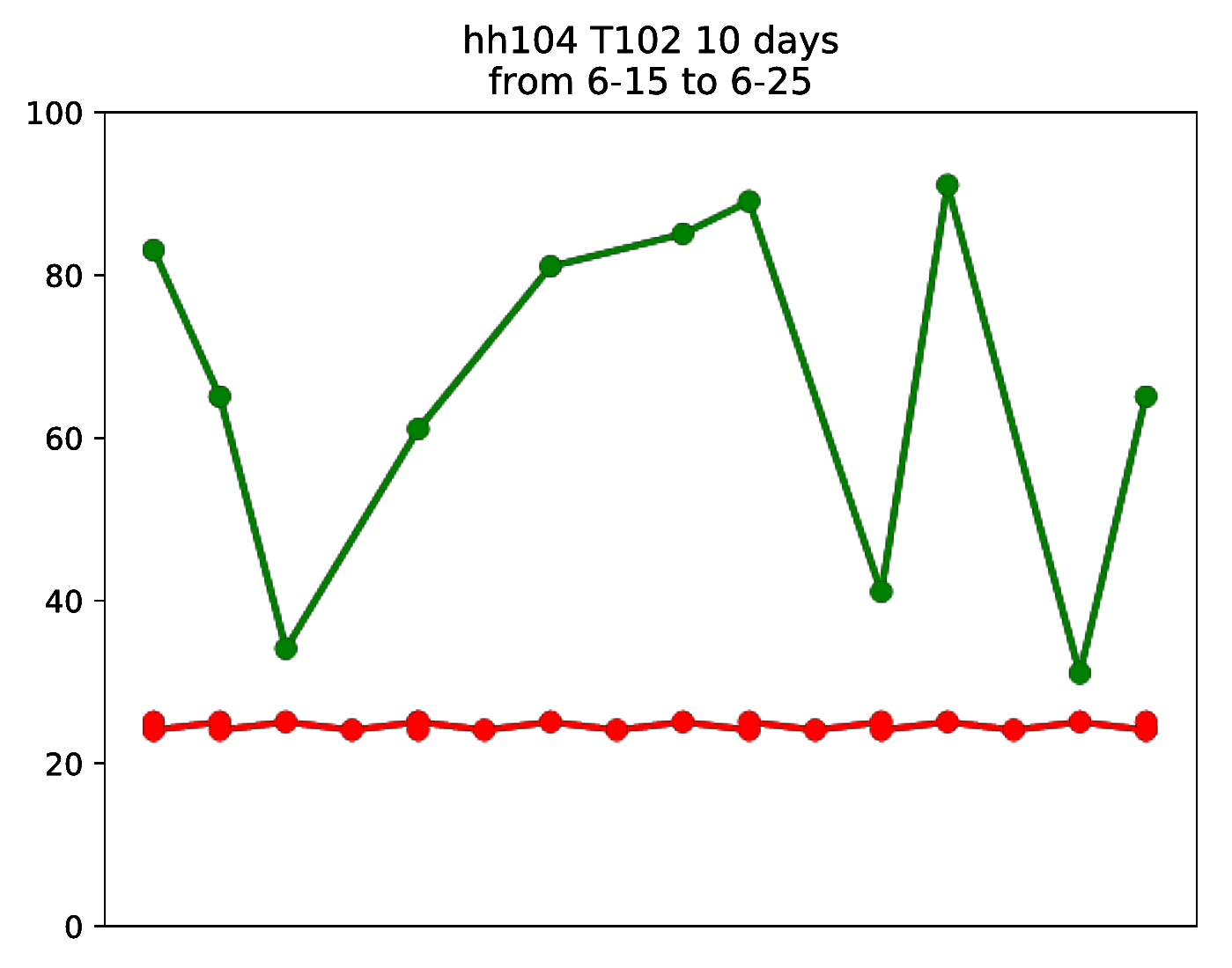}
	}
	\subfigure[] {
		\includegraphics[width=0.45\columnwidth]{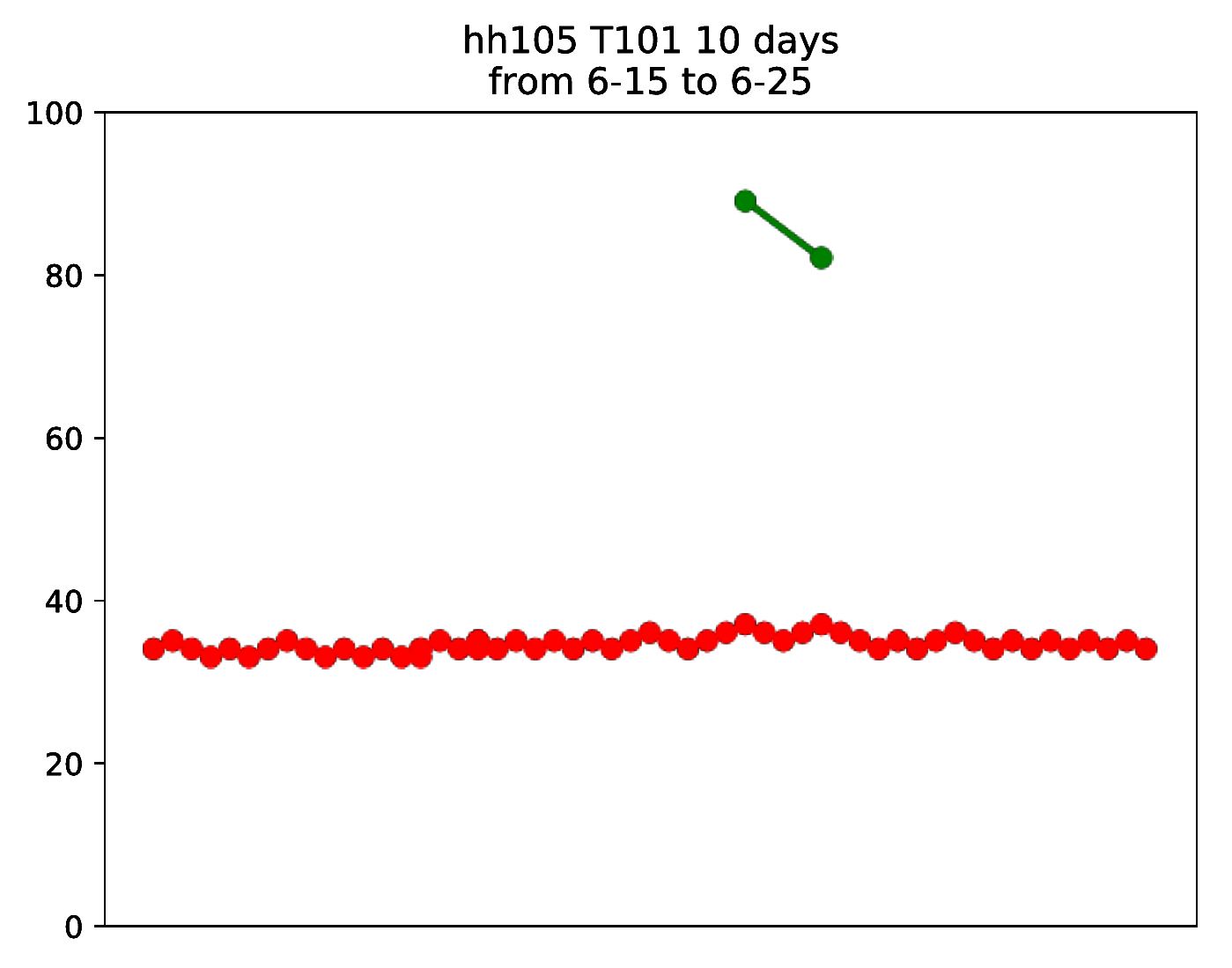}
	}
	\subfigure[] {
		\includegraphics[width=0.45\columnwidth]{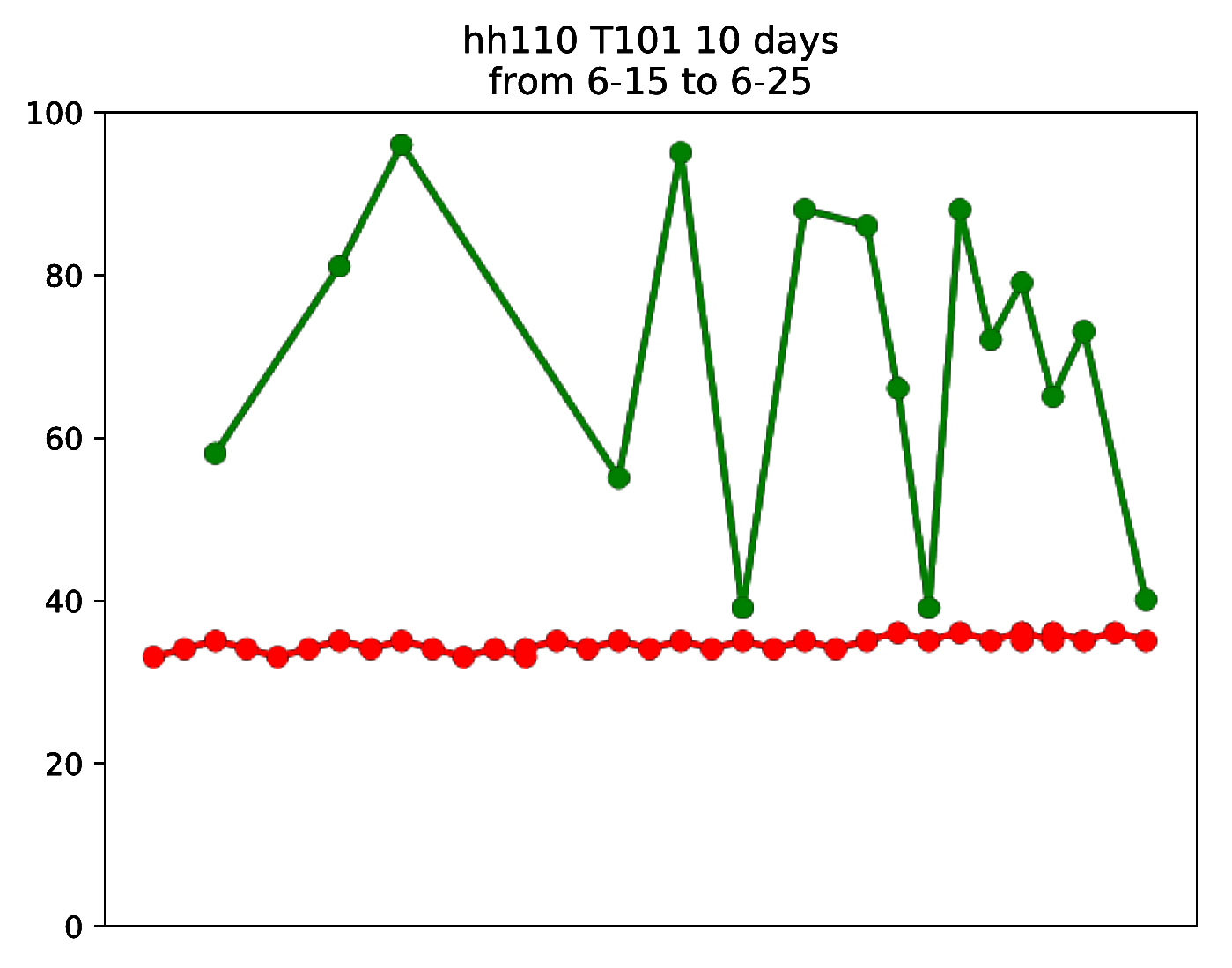}
	}
	\subfigure[] {
		\includegraphics[width=0.45\columnwidth]{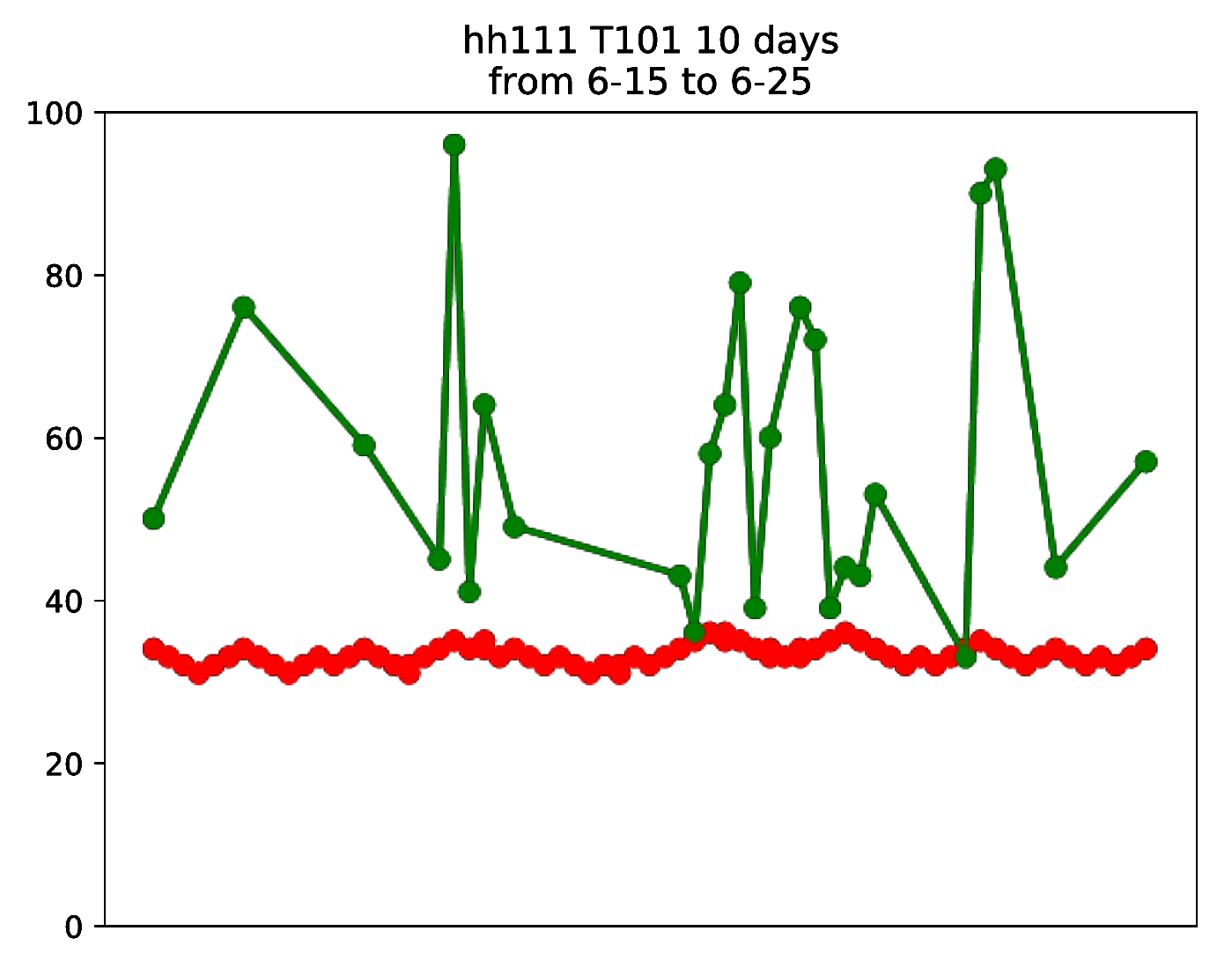}
	}
	\caption{Comparison the numeric values of eight temperature sensors from different datasets that are uploaded to two datasets.}
	\label{F_Eight_comparing_numeric_values}
\end{figure}

\subsection{Comparing the Numeric Records}

For numeric records, \FF~filters redundant events and randomizes their values. We believe that these fluctuations can also reflect a user's behavior patterns. Then, IFTTT cannot track the fluctuation of the temperature, humidity, and illuminance in a living environment. Fig.~\ref{F_Eight_comparing_numeric_values} shows the comparison of the records from selected temperature sensors that are either filtered by \FF~or not. The red line plots the records accessible to IFTTT, while the green line plots records that are processed by \FF. This proves that after filtering by \FF, IFTTT can only get a few records, none of which are accurate (only the first 10 days' records are shown).

\section{\FF: the fuzzing component}
\label{fuzz}

In this section, we will introduce the fuzzing component of \FF~that eliminates the statistical characters of the events that uploaded to IFTTT or the other TA platforms.

\subsection{Behavior Pattern of A Trigger-Action User}

For the Trigger-Action model, the 3rd-party platforms can profile a user's behavior pattern through raw data. As a typical representation of 3rd-party platforms, IFTTT can obtain plenty of information just from the time of the event records without any sophisticated mechanisms. For example, the daily routine (leaving and returning home), physical health status (activity frequency in the bathroom when it is late at night), and living environment, \etc, by present sensors, motion sensors, and temperature/humidity/illuminance sensors, respectively. So, we can treat a user's behavior pattern as the random variable of events at different time intervals of a day. After observing a user for several days, we can get the random variable $X = X(h)$ to denote the average amount of the events from a specific devices in different intervals (\eg~if $h$ denotes different hours) of a day. If the observation time is long enough, each $X = X(h)$ will approach a static value. Then, each user will own a unique vector of $[X(1), \cdots, X(24)]$. In other words, we can treat the vector $[X(1), \cdots, X(24)]$ as a user's behavior pattern after sufficient observation. We use $U$ to denote this vector, then $[X(1), \cdots, X(24)]$ can also be denoted as $[u_1, \cdots, u_{24}]$.

After filtering redundant events, we know that the amount of event records uploaded to IFTTT has been significantly reduced. Therefore, to the IFTTT platformsome, some information relative to the user has been lost (by intercepting the redundant event records) and this becomes an obstacle for IFTTT to find some behavior patterns of a user using data mining or statistics approaches. But after filtering the records, the event records that have to be uploaded to IFTTT or other 3rd-party platforms can still reveal a user's behavior patterns. 

\subsection{Privacy Leakage via the Filtered Event Records}

After filtering, each single record that is uploaded to IFTTT contains information of when it happened (time of day, day of week, weekend or workday, and so on). For instance, if a user deployed several sensors and set an app of ``If any new motion is detected by the $motion\_sensor$, then turn on the $light$.'' After filtering, IFTTT and other 3rd-party platforms will obtain fewer records because the events from other sensors will all be intercepted, and the events from this $motion\_sensor$ may be intercepted with a probability. We can compare the original and filtered event records of the ``turn on the light'' app after several days. From the original event records, we can obtain the vector $U$. At the same time, we can also obtain $F$ as another vector that contains the statistical character of the filtered event records. For the $i$-th elements in $U$ and $F$, which denote the mean amount of the events in the $i$-the hour of a day, $f_i$ is less than $u_i$ since the events in the $i$-th hour may be intercepted by chance. Therefore, $F = [f_1, \cdots, f_{24}] = [u_1 \times p_1, \cdots, u_{24} \times p_{24}]$, and $P = [p_1, \cdots, p_{24}]$ is the vector that denotes the probability of intercepting an event that is to be uploaded to IFTTT in each hour.

In most cases, $p_i$ will be a relatively stable value in $P$. Then there will be an extreme high correlation coefficient between $[u_1, \cdots, u_{24}]$ and $[f_1, \cdots, f_{24}]$. This means that, just filtering the original event records is not adequate for preventing IFTTT or other 3rd-party platforms to investigate a user's behavior patterns.

\begin{figure}[tbp] \centering
	\includegraphics[width=0.95\columnwidth]{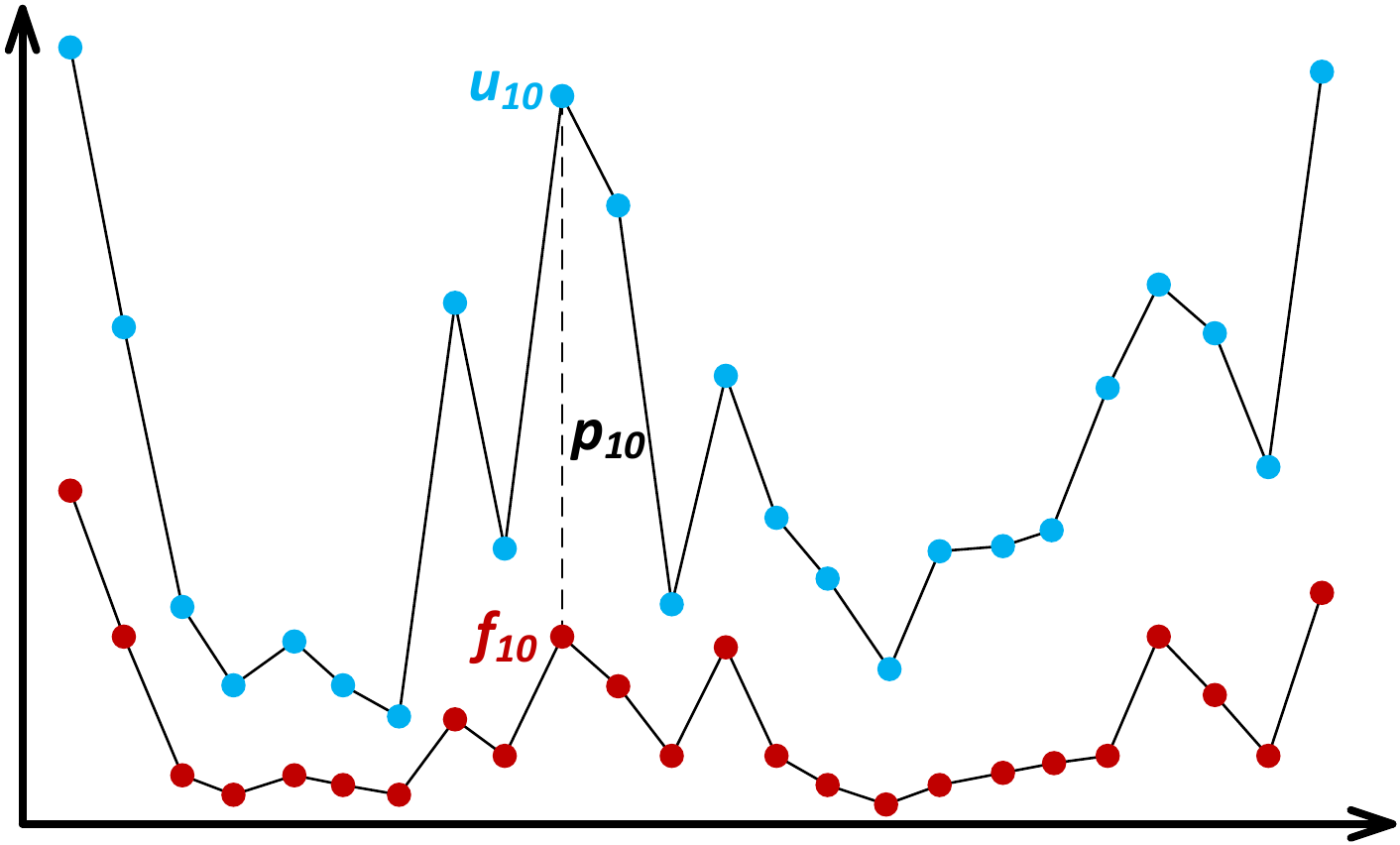}
	\caption{Filtered records shows highly relative to original records.}
	\label{F_Sample}
\end{figure}

We prove this by monitoring a user living in a single apartment for 10 days. This user deployed several sensors and actuators, including a contact sensor, two motion sensors, and five lights. This user set the app ``If any new motion is detected by the $motion\_sensor$, then turn on the $light$'' to automatically turn on the light in his restroom. After 10 days observing, we obtain $U$ and $F$, which are shown in Fig~\ref{F_Sample}.

We can observe that although the sum of all items in $F$ is much less than $U$, the statistical character of $U$ was quite similar to $F$. The behavior patterns inferred from $U$ can be also obtained from $F$. This proves that filtering the original event records still can expose information about a user's behavior patterns after a short observation period.

\subsection{Fuzzing component that conceals the statistical charater of events}

\subsubsection{Main idea} Due to the stateless aspect of Trigger-Action IoT apps, IFTTT and other 3rd-party platforms that run TA apps are only responsible for replying if the current event comes from the smart home. When a new event arrives, all IFTTT needs to do is send back the corresponding command message (if any) to the smart home platform. Our goal is to mix some pseudo-events into the filtered event records. As the 3rd-party platform can not distinguish between genuine and pseudo events, we can use this method to ``fuzz'' the filtered dataset and consequently make IFTTT and other 3rd-party platforms unable to reveal the statistical character of the user's actual behavior pattern. If the dataset obtained by IFTTT shows no correlation to the filtered dataset, IFTTT will learn nothing about the filtered event dataset. The same applies for other 3rd-party platforms.

\subsubsection{The fuzzing workflow} We design the fuzzing feature as an additional component that works between the smart home platform and IFTTT. The fuzzing component of \FF~can preserve the regular functionality of the apps running on IFTTT and other 3rd-party platforms, and conceals the user's behavior patterns. This component works between a smart home platform $S$ and several 3rd-party platforms. For a specific 3rd-party platform, for example, IFTTT, we use $T$ to denote it. For the current communication between $S$ and $T$, there are many users who trigger their sensors and yield events each second, then $S$ will pack up these events and send them to $T$. Finally, $S$ waits for the command messages from $T$. But the fuzzing component of \FF~modifies this procedure and works as follows.

\begin{enumerate}
	\item In each second, before sending the ``real'' events to $T$, $S$ will generate a number of pseudo-events by \texttt{Fuzz()} and will pack up the real-events and pseudo-events. \texttt{Fuzz()} can generate pseudo-events corresponding to any devices of all the users.
	\item To differentiate the real and pseudo events, $S$ will maintain a list in which each item is the state of a device. Each item is a tuple with the format of [\texttt{time stamp}, \texttt{user id}, \texttt{device id}, \texttt{event value}, \texttt{pseudo-label}].
	\item Next, after receiving the event list from $S$, $T$ will send back a list of commands corresponding the different apps that triggered by the event records in the event list. As $T$ can not distinguish whether the items in the event list are true or not, it will reply to the whole event list with a command list. The format of the items in the command list is [\texttt{time stamp}, \texttt{user id}, \texttt{device id}, \texttt{command}].
	\item Finally, $S$ receives the list of the commands and only retains the commands that correspond to actual events by matching the \texttt{time stamp}, \texttt{user id}, and \texttt{device id}, as well as checking the \texttt{pseudo-label}. By checking the \texttt{pseudo-label}, $S$ will decide whether to deliver the command to the device or just discard it.
\end{enumerate}

By adopting a proper \texttt{Fuzz()}, the actual and pseudo events are indistinguishable to any 3rd-party platforms, including IFTTT. Therefore, the statistical character of the actual events will be harder or even impossible to be extracted by IFTTT.

\subsection{The Pseudo-events Generator}

As the pseudo-events generator, \texttt{Fuzz()} should achieve two goals at the same time: the pseudo events can efficiently fuzz IFTTT, and can constrain the extra overhead of smart homes and IFTTT. For a specific user, the most ideal \texttt{Fuzz()} is one that can get rid of all the statistical characters for all of his devices.

\subsubsection{Generating pseudo events in an intuitive way} To any device of a specific user, an intuitive way of designing \texttt{Fuzz()} is generating its pseudo events with an arbitrary possibility that is irrelevant to the behavior patterns of the user. However, by the law of Large Numbers, we know that after a long enough period of time, the amount of pseudo events will form a uniform distribution for each time interval. Then the distribution of the event records that are obtained by the 3rd-party platform will be exactly identical to the event records without any pseudo events. Therefore, \texttt{Fuzz()} implemented in this way will be meaningless for protecting a user's behavior patterns.

\subsubsection{Dynamic \texttt{Fuzz()}} We designed the \texttt{Fuzz()} algorithm with a feature that allows it to adjust itself by generating pseudo events according to the actual behavior patterns of a user. By observing the user's behavior pattern during several recent days, \texttt{Fuzz()} can generate pseudo events by adjusting itself so that the amount of event records obtained by the 3rd-party platform will remain in a static pattern and carry no information about a user's behavior patterns. To achieve this, \texttt{Fuzz()} sets a target distribution $D$. Then, \texttt{Fuzz()} will dynamically adjust itself to generate the pseudo events to make sure the event records obtained by the 3rd party platform are always distributed as $D$. This ensures that the 3rd party learns nothing from this static distribution. 

\subsubsection{The steps of dynamic \texttt{Fuzz()}} In order to make the sum of actual and pseudo events equal to a relatively stable value, the essential dynamic of \texttt{Fuzz()} is first establishing $D$, then adjusting the probability of generating pseudo events as time goes on. We denote the distribution of the events after filtering as $F$. $F$ is quite similar to the original behavior pattern vector $U$, as we mentioned above. Then the steps of dynamic \texttt{Fuzz()} are as follows.
\begin{enumerate}
	\item Monitor a user for $p$ days to obtain/update the behavior pattern vector $F$.
	\item Establish the target vector $D$ with $n$ elements (\eg~$n$ = 24). These $n$ elements correspond to $n$ portions of a whole day and can have arbitrary distributions for various user behavior patterns, but the max of $D$ should equal the max of $F$. For each $d_i$ in $D$:
	\begin{enumerate}
		\item If elements of $D$ have a uniform distribution, then $d_1$ = $d_2$ = $\cdots$ = $d_n = max(f_i)$.
		\item If elements of $D$ have a Gaussian distribution, then $\mu$ (the expected value of $D$) is equals $max(f_i)$. According to the 3-$\sigma$ rule, $\sigma$ should be equal or greater than $n/6$.
	\end{enumerate}
	\item Get another $n$-element vector $Y$. For each element $y_i$ in $Y$:
	\begin{enumerate}
		\item If $d_i \geq f_i$, $y_i = d_i - f_i$;
		\item If $d_i < f_i$, $y_i = 0$.
	\end{enumerate}
	\item We assume that in each of the $n$ portions, the smart home platform will send the event records for $m$ times. Then, for each time the smart home platform packs the records:
	\begin{enumerate}
		\item If there is an actual event, then \texttt{Fuzz()} does nothing and the smart home will send the actual event to the 3rd-party platform;
		\item If there is no actual event, then \texttt{Fuzz()} will generate a pseudo event by a probability of $y_i/m$.
	\end{enumerate}
	\item Adjust $F$ according to the filtered event records during the past $p$ days and repeat 2 \textendash~4. 
\end{enumerate}

\subsection{Two \texttt{Fuzz()} proposals}

We propose two varieties of \texttt{Fuzz()}. For each \texttt{Fuzz()}, the effects and overheads are different.

\subsubsection{The ideal \texttt{Fuzz()}} 
\begin{figure}[tbp] \centering
	\includegraphics[width=0.95\columnwidth]{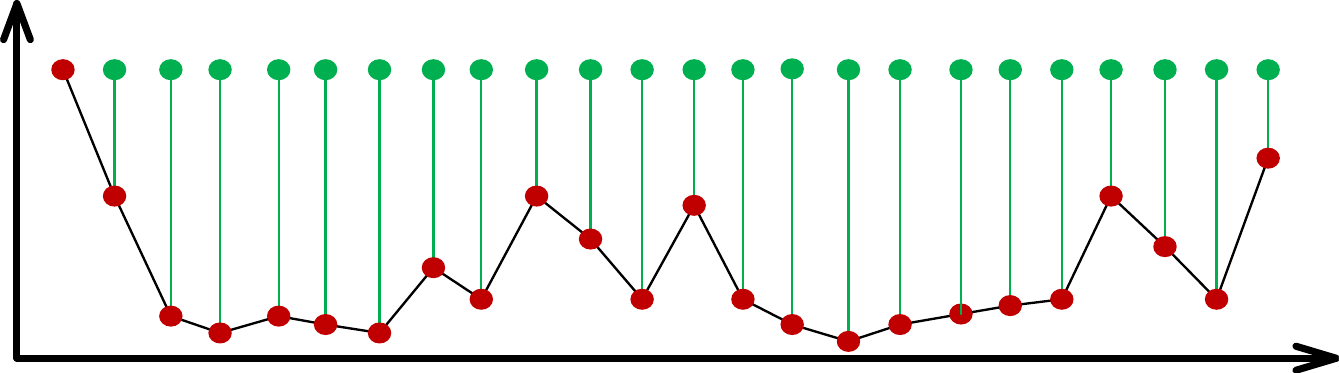}
	\caption{\texttt{Mask()} with ideal fuzzification.}
	\label{Ideal_fuzzification}
\end{figure}

As we know, an array containing fixed elements shows no correlations to any other arrays. If we make the elements in $D$ all relatively stable, then it will be more difficult for a 3rd-party platform to infer the original distribution of a user's event records, namely the characters of $F$ or $U$. In an ideal scenario, all elements in $D$ will be identical values and the correlation coefficient between $D$ and $F$ will be 0. In this case, the elements of $D$ have a uniform distribution. In order to cover the peak of $F$, we let $d_1$ = $d_2$ = $\cdots$ = $d_n = max(f_i)$. This can be illustrated in Fig~\ref{Ideal_fuzzification}.

\subsubsection{\texttt{Fuzz()} with a Gaussian distribution} If $D$ is a vector that does not contain fixed but variable elements, it can also hide the characteristics of $F$ to prevent the 3rd-party platforms from learning something from $F$. This can be done by making $D$ a static vector. Then, maybe the relationship coefficient between $D$ and $F$ is greater than it is between $D$ with uniform a distribution and $F$, but IFTTT still can not infer many characters because various $F$ vectors can be masked by an identical $D$ vector. 

As we can see, for the $i$-th element in $D$, the overhead is $y_i = d_i - f_i$. For all the elements in $Y$, $y_i$ can be reduced if $f_i$ is a relatively small value because $d_i$ can also be set as a small value (but can not be less than $f_i$).

\begin{figure}[tbp] \centering
	\includegraphics[width=0.95\columnwidth]{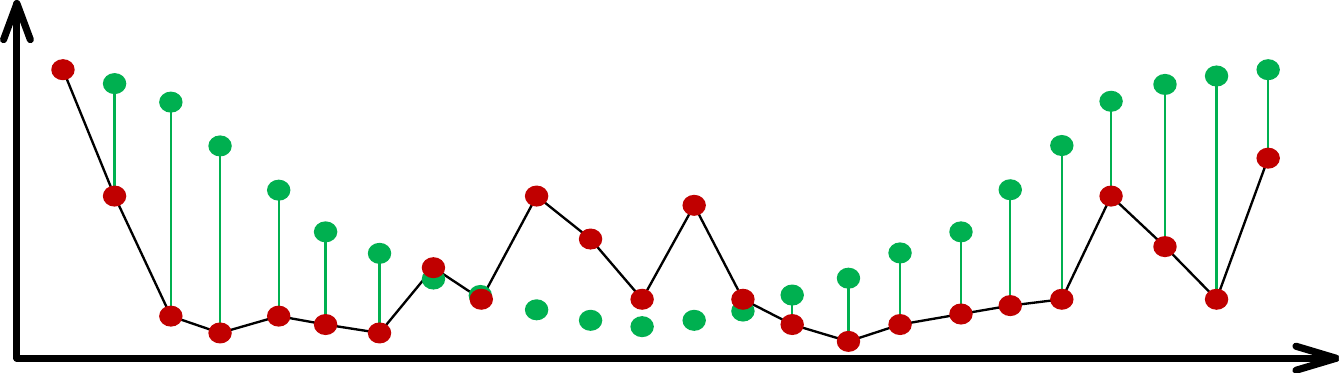}
	\caption{\texttt{Mask()} with a Gaussian distribution.}
	\label{Gaussian_distribution}
\end{figure}

Therefore, we can reduce the overhead by modifying the elements in $D$ to make them have a Gaussian distribution. Then $\mu$ (the expected value of $D$) equals $max(f_i)$. According to the 3-$\sigma$ rule, $\sigma$ should be equal or greater than $n/6$. Besides the peak element, all other elements are less than $max(f_i)$; this can make $D$ a static vector and the sum of all the elements less than it is with a uniform distribution. This can be illustrated in Fig~\ref{Gaussian_distribution}.

\section{Evaluation}  

In this section, we evaluate \texttt{Fuzz()} by: 1) the amount of pseudo events generated by \texttt{Fuzz()} and 2) how some representative machine learning algorithms can figure out the characters of a vector.

\subsection{Design of the Experiments for Evaluation}

The main idea of our experiments is evaluating how much personal profiling information is still contained in the behavior pattern vectors. We used two prevalant machine learning algorithms, KNN and SVM, to compare the success rate of identifying a user's pattern vector before and after mixing pseudo events. Additionally, we also compare the amount of the pseudo events by different kinds of \texttt{Fuzz()}.

\subsection{Dataset for Training and Testing}

For both the actual and pseudo event records, we all can obtain the corresponding behavior pattern vectors by days. After 100 days observing two volunteers who lived in two single apartments with identical layouts, we obtained 100 vectors of filtered event records for each of them. We encouraged them to keep their normal behavior patterns, and two actual vectors between the two volunteers had a correlation coefficient of nearly 0.93.

We used all the vectors from each volunteer to obtain a 200-vector dataset, and then tested KNN and SVM. We split the dataset into training (70\%) and testing (30\%) portions. The results showed that though the correlation coefficient between the behavior pattern vectors of the two volunteers is as high as 0.93, after more than 10 experiments, the correct rates of distinguishing a behavior pattern vector were at least 98\%. This proved the accuracy and high performance of the two machine learning algorithms.

\subsection{Evaluating the Overhead and Masking Effect}

We evalutated the overhead and fuzzing effect of the two \texttt{Fuzz()} schemes, corresponding to two privacy protection levels. As we can see from Fig~\ref{Ideal_fuzzification} and Fig~\ref{Gaussian_distribution}, the amount of pseudo-records inserted by the ideal fuzzification is more than that of a Gaussian distribution. But we can also see that in a Gaussian distribution, some elements of a vector cannot be masked, and the result is kept as a static character. This decreases the security of masking with Gaussian distribution and also lowers the overhead compared to the ideal fuzzification. After 100 days, there were 1772 and 2204 event records from each of the two users. The comparison results can be shown in Fig.~\ref{Evaluation_result}

\subsubsection{Overhead} We evaluated how many pseudo records should be inserted into these two datasets, and after 10 experiments, the ratio of the amount of records after masking and filtering is shown in Fig.~\ref{sub_a}. As we can see, the ideal fuzzification needs more pseudo items to mask the character of the filtered records.

\subsubsection{Correlation between the masked records and filtered records} If the \texttt{Fuzz()} adopts the ideal fuzzing algorithm, the correlation coefficient between the masked records and filtered records is roughly around 0, which means that the masked records will hardly reveal anything about the filtered records. But when \texttt{Fuzz()} adopts a Gaussian distribution on the masked records, this result will increase to 0.68. Fig.~\ref{sub_b} shows the results of 10 experiments.

\subsubsection{Success rate of KNN and SVM} 
\begin{figure*}[tbp] \centering
	\subfigure[The ratio of the amount of records after masking and the amount of the records after filtering.] {
		\includegraphics[width=0.31\textwidth]{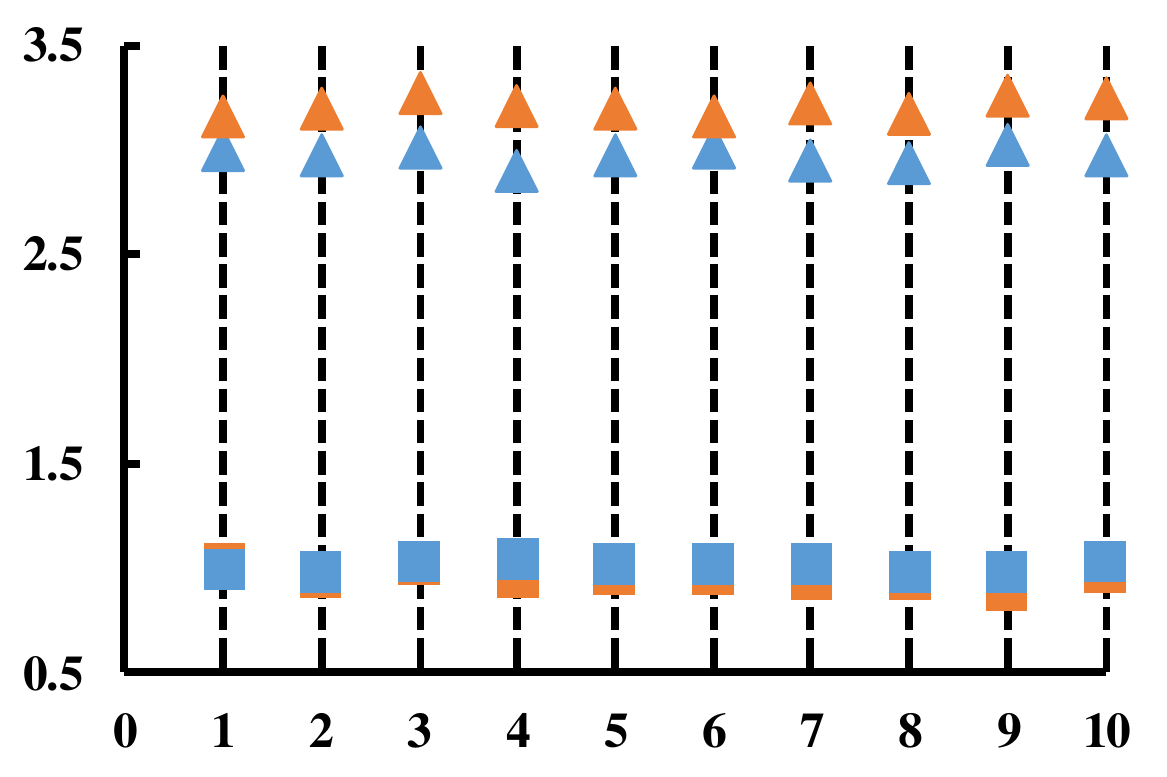}
		\label{sub_a}
	}
	\subfigure[The correlation coefficient between the records after masking and after filtering.] {
		\includegraphics[width=0.31\textwidth]{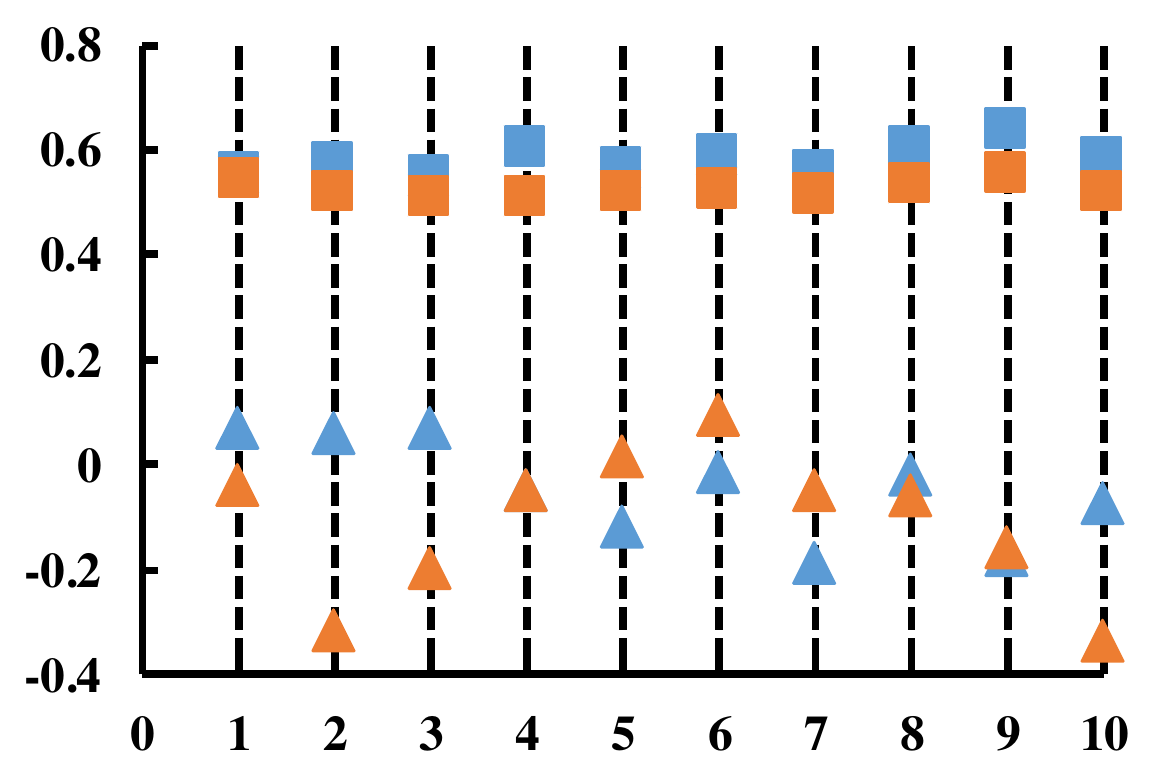}
		\label{sub_b}
	}
	\subfigure[The success rate of KNN and SVM.] {
		\includegraphics[width=0.31\textwidth]{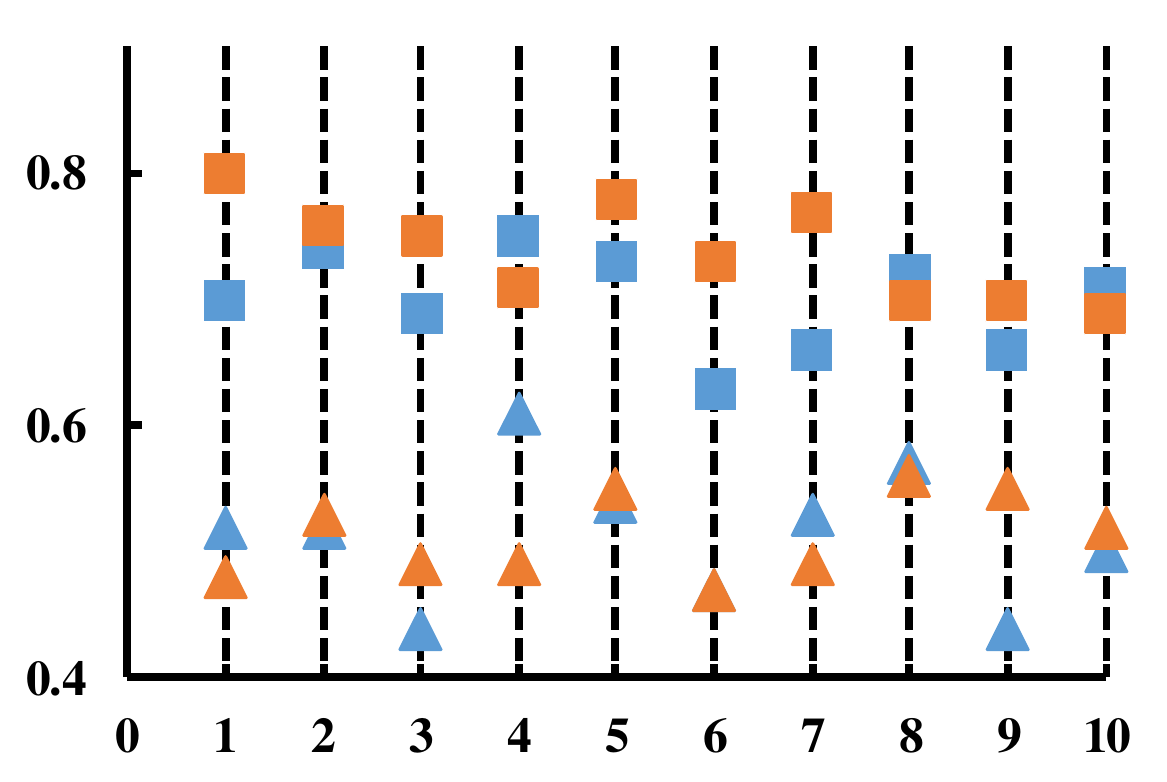}
		\label{sub_c}
	}
	\caption{Evaluation result ($\blacktriangle$: Ideal fuzzification; $\blacksquare$: Gaussian distribution).}
	\label{Evaluation_result}
\end{figure*}

As we collected the event records from two users, when \texttt{Fuzz()} showed a stronger effect to eliminate the statistical character of the users, the successful rate of distinguishing the two users by a vector would be closer to 0.5. Our experiment verified the effectiveness of \texttt{Fuzz()} with ideal fuzzing and Gaussian distribution. As shown in Fig.~\ref{sub_c}, when \texttt{Fuzz()} adopts a Gaussian distribution, the success rate of KNN and SVM is about 0.7. Considering the fact that the success rate will be 0.5 if \texttt{Fuzz()} just randomly determines the group of a vector, a success rate of 0.7 means these two machine learning algorithms were ``guessing randomly'' in most vectors. The ideal \texttt{Fuzz()} has a success rate close to 0.5, proving its efficiency. This is because it means that no personal behavioral information is leaked to IFTTT and other 3rd-party platforms. 

\section{Conclusion}

Security and privacy will be of ongoing interest, especially in the smart home environment. In this paper, we studied how IFTTT monitors its smart home users in a hidden way and proposed \FF~for preventing this privacy leakage by: filtering the redundant event records that are uploaded to IFTTT and concealing the statistical characters of the events to fuzz IFTTT. We use IFTTT as the representative target, but we believe that \FF~can also be applied to other integrated 3rd-party services with the same architecture.


\begin{thebibliography}{00}
		\providecommand{\url}[1]{#1}
		\csname url@samestyle\endcsname
		\providecommand{\newblock}{\relax}
		\providecommand{\bibinfo}[2]{#2}
		\providecommand{\BIBentrySTDinterwordspacing}{\spaceskip=0pt\relax}
		\providecommand{\BIBentryALTinterwordstretchfactor}{4}
		\providecommand{\BIBentryALTinterwordspacing}{\spaceskip=\fontdimen2\font plus
			\BIBentryALTinterwordstretchfactor\fontdimen3\font minus
			\fontdimen4\font\relax}
		\providecommand{\BIBforeignlanguage}[2]{{%
				\expandafter\ifx\csname l@#1\endcsname\relax
				\typeout{** WARNING: IEEEtran.bst: No hyphenation pattern has been}%
				\typeout{** loaded for the language `#1'. Using the pattern for}%
				\typeout{** the default language instead.}%
				\else
				\language=\csname l@#1\endcsname
				\fi
				#2}}
		\providecommand{\BIBdecl}{\relax}
		\BIBdecl
		
		\bibitem{SmartThi62:online}
		Samsung, ``Smartthings. {A}dd a little smartness to your things.''
		\url{https://www.smartthings.com/}, (Accessed on July 2018).
		
		\bibitem{WeaveNes79:online}
		Google, ``Weave | {N}est,'' \url{https://nest.com/weave/}, (Accessed on July
		2018).
		
		\bibitem{iOSHomeA95:online}
		Apple, ``i{OS} - {H}ome - {A}pple,'' \url{https://www.apple.com/ios/home/},
		(Accessed on July 2018).
		
		\bibitem{Fernandes2016}
		\BIBentryALTinterwordspacing
		E.~Fernandes, J.~Jung, and A.~Prakash, ``{Security Analysis of Emerging Smart
			Home Applications},'' \emph{2016 IEEE Symposium on Security and Privacy
			(SP)}, pp. 636--654, 2016. [Online]. Available:
		\url{http://ieeexplore.ieee.org/document/7546527/}
		\BIBentrySTDinterwordspacing
		
		\bibitem{martin_finnegan_2018}
		\BIBentryALTinterwordspacing
		J.~A. Martin and M.~Finnegan, ``What is {IFTTT}? {H}ow to use {I}f {T}his,
		{T}hen {T}hat services,'' Feb 2018. [Online]. Available:
		\url{https://www.computerworld.com/article/3239304/mobile-wireless/what-is-ifttt-how-to-use-if-this-then-that-services.html}
		\BIBentrySTDinterwordspacing
		
		\bibitem{smartthingscommunity_2018}
		\BIBentryALTinterwordspacing
		SmartThingsCommunity, ``Smartthingscommunity · github.'' [Online]. Available:
		\url{https://github.com/SmartThingsCommunity}
		\BIBentrySTDinterwordspacing
		
		\bibitem{du2005designing}
		X.~Du and F.~Lin, ``Designing efficient routing protocol for heterogeneous
		sensor networks,'' in \emph{Performance, Computing, and Communications
			Conference, 2005. IPCCC 2005. 24th IEEE International}.\hskip 1em plus 0.5em
		minus 0.4em\relax IEEE, 2005, pp. 51--58.
		
		\bibitem{du2006adaptive}
		X.~Du and D.~Wu, ``Adaptive cell relay routing protocol for mobile ad hoc
		networks,'' \emph{IEEE Transactions on Vehicular Technology}, vol.~55, no.~1,
		pp. 278--285, 2006.
		
		\bibitem{du2004qos}
		X.~Du, ``Qos routing based on multi-class nodes for mobile ad hoc networks,''
		\emph{Ad Hoc Networks}, vol.~2, no.~3, pp. 241--254, 2004.
		
		\bibitem{mandala2008load}
		D.~Mandala, X.~Du, F.~Dai, and C.~You, ``Load balance and energy efficient data
		gathering in wireless sensor networks,'' \emph{Wireless Communications and
			Mobile Computing}, vol.~8, no.~5, pp. 645--659, 2008.
		
		\bibitem{Islam2012}
		\BIBentryALTinterwordspacing
		K.~Islam, W.~Shen, and X.~Wang, ``{Security and privacy considerations for
			Wireless Sensor Networks in smart home environments},'' \emph{Proceedings of
			the 2012 IEEE 16th International Conference on Computer Supported Cooperative
			Work in Design (CSCWD)}, pp. 626--633, 2012. [Online]. Available:
		\url{http://ieeexplore.ieee.org/document/6221884/}
		\BIBentrySTDinterwordspacing
		
		\bibitem{Lee2014}
		C.~Lee, L.~Zappaterra, K.~Choi, and H.-a. Choi, ``{Securing Smart Home :
			Technologies , Security Challenges , and Security Requirements},'' pp.
		67--72, 2014.
		
		\bibitem{Lin2016}
		\BIBentryALTinterwordspacing
		H.~Lin and N.~Bergmann, ``{IoT Privacy and Security Challenges for Smart Home
			Environments},'' \emph{Information}, vol.~7, no.~3, p.~44, 2016. [Online].
		Available: \url{http://www.mdpi.com/2078-2489/7/3/44}
		\BIBentrySTDinterwordspacing
		
		\bibitem{RisteskaStojkoska2017}
		\BIBentryALTinterwordspacing
		B.~L. {Risteska Stojkoska} and K.~V. Trivodaliev, ``{A review of Internet of
			Things for smart home: Challenges and solutions},'' \emph{Journal of Cleaner
			Production}, vol. 140, pp. 1454--1464, 2017. [Online]. Available:
		\url{http://dx.doi.org/10.1016/j.jclepro.2016.10.006}
		\BIBentrySTDinterwordspacing
		
		\bibitem{cai2018private}
		Z.~Cai and X.~Zheng, ``A private and efficient mechanism for data uploading in
		smart cyber-physical systems,'' \emph{IEEE Transactions on Network Science
			and Engineering}, 2018.
		
		\bibitem{hei2013pipac}
		X.~Hei, X.~Du, S.~Lin, and I.~Lee, ``Pipac: Patient infusion pattern based
		access control scheme for wireless insulin pump system.'' in \emph{INFOCOM},
		2013, pp. 3030--3038.
		
		\bibitem{Cheng2017}
		\BIBentryALTinterwordspacing
		Y.~Cheng, X.~Fu, X.~Du, B.~Luo, and M.~Guizani, ``{A lightweight live memory
			forensic approach based on hardware virtualization},'' \emph{Information
			Sciences}, vol. 379, pp. 23--41, 2017. [Online]. Available:
		\url{http://dx.doi.org/10.1016/j.ins.2016.07.019}
		\BIBentrySTDinterwordspacing
		
		\bibitem{Du2007}
		X.~Du, Y.~Xiao, M.~Guizani, and H.~H. Chen, ``{An effective key management
			scheme for heterogeneous sensor networks},'' \emph{Ad Hoc Networks}, vol.~5,
		no.~1, pp. 24--34, 2007.
		
		\bibitem{Sivaraman2015}
		V.~Sivaraman, H.~H. Gharakheili, A.~Vishwanath, R.~Boreli, and O.~Mehani,
		``{Network-level security and privacy control for smart-home IoT devices},''
		\emph{2015 IEEE 11th International Conference on Wireless and Mobile
			Computing, Networking and Communications, WiMob 2015}, pp. 163--167, 2015.
		
		\bibitem{Ali2017}
		W.~Ali, G.~Dustgeer, M.~Awais, and M.~A. Shah, ``{IoT based smart home:
			Security challenges, security requirements and solutions},''
		\emph{International Conference on Automation and Computing (ICAC)}, no.
		September, pp. 1--6, 2017.
		
		\bibitem{Jose2017}
		A.~C. Jose and R.~Malekian, ``{Improving Smart Home Security: Integrating
			Logical Sensing into Smart Home},'' \emph{IEEE Sensors Journal}, vol.~17,
		no.~13, pp. 4269--4286, 2017.
		
		\bibitem{Dorri2017}
		\BIBentryALTinterwordspacing
		A.~Dorri, S.~S. Kanhere, R.~Jurdak, and P.~Gauravaram, ``{Blockchain for IoT
			security and privacy: The case study of a smart home},'' \emph{2017 IEEE
			International Conference on Pervasive Computing and Communications Workshops
			(PerCom Workshops)}, pp. 618--623, 2017. [Online]. Available:
		\url{http://ieeexplore.ieee.org/document/7917634/}
		\BIBentrySTDinterwordspacing
		
		\bibitem{du2009transactions}
		X.~Du, M.~Guizani, Y.~Xiao, and H.-H. Chen, ``Transactions papers a
		routing-driven elliptic curve cryptography based key management scheme for
		heterogeneous sensor networks,'' \emph{IEEE Transactions on Wireless
			Communications}, vol.~8, no.~3, pp. 1223--1229, 2009.
		
		\bibitem{Lee2017}
		\BIBentryALTinterwordspacing
		S.~Lee, J.~Choi, J.~Kim, B.~Cho, S.~Lee, H.~Kim, and J.~Kim, ``{FACT:
			Functionality-centric Access Control System for IoT Programming
			Frameworks},'' \emph{Proceedings of the 22nd ACM on Symposium on Access
			Control Models and Technologies}, pp. 43--54, 2017. [Online]. Available:
		\url{http://doi.acm.org/10.1145/3078861.3078864}
		\BIBentrySTDinterwordspacing
		
		\bibitem{Tian2017}
		\BIBentryALTinterwordspacing
		Y.~Tian, N.~Zhang, Y.-H. Lin, X.~Wang, B.~Ur, X.~Guo, and P.~Tague,
		``{SmartAuth: User-Centered Authorization for the Internet of Things},''
		\emph{Usenix}, pp. 361--378, 2017. [Online]. Available:
		\url{https://www.usenix.org/conference/usenixsecurity17/technical-sessions/presentation/tian}
		\BIBentrySTDinterwordspacing
		
		\bibitem{Jia2017}
		\BIBentryALTinterwordspacing
		Y.~J. Jia, Q.~A. Chen, S.~Wang, A.~Rahmati, E.~Fernandes, Z.~M. Mao, and
		A.~Prakash, ``{ContexIoT: Towards Providing Contextual Integrity to Appified
			IoT Platforms},'' \emph{Proceedings 2017 Network and Distributed System
			Security Symposium}, no. March, 2017. [Online]. Available:
		\url{https://www.ndss-symposium.org/ndss2017/ndss-2017-programme/contexlot-towards-providing-contextual-integrity-appified-iot-platforms/}
		\BIBentrySTDinterwordspacing
		
		\bibitem{Celik2018}
		\BIBentryALTinterwordspacing
		Z.~B. Celik, L.~Babun, A.~K. Sikder, H.~Aksu, G.~Tan, P.~McDaniel, and A.~S.
		Uluagac, ``{Sensitive Information Tracking in Commodity IoT},'' pp. 1--17,
		2018. [Online]. Available: \url{http://arxiv.org/abs/1802.08307}
		\BIBentrySTDinterwordspacing
		
		\bibitem{Geneiatakis2017}
		\BIBentryALTinterwordspacing
		D.~Geneiatakis, I.~Kounelis, R.~Neisse, I.~Nai-Fovino, G.~Steri, and
		G.~Baldini, ``{Security and privacy issues for an IoT based smart home},''
		\emph{2017 40th International Convention on Information and Communication
			Technology, Electronics and Microelectronics (MIPRO)}, pp. 1292--1297, 2017.
		[Online]. Available: \url{http://ieeexplore.ieee.org/document/7973622/}
		\BIBentrySTDinterwordspacing
		
		\bibitem{Wu2016}
		L.~Wu, X.~Du, and J.~Wu, ``{Effective Defense Schemes for Phishing Attacks on
			Mobile Computing Platforms},'' \emph{IEEE Transactions on Vehicular
			Technology}, vol.~65, no.~8, pp. 6678--6691, 2016.
		
		\bibitem{liang2018deep}
		Y.~Liang, Z.~Cai, J.~Yu, Q.~Han, and Y.~Li, ``Deep learning based inference of
		private information using embedded sensors in smart devices,'' \emph{IEEE
			Network}, vol.~32, no.~4, pp. 8--14, 2018.
		
		\bibitem{xiao2007survey}
		Y.~Xiao, V.~K. Rayi, B.~Sun, X.~Du, F.~Hu, and M.~Galloway, ``A survey of key
		management schemes in wireless sensor networks,'' \emph{Computer
			communications}, vol.~30, no. 11-12, pp. 2314--2341, 2007.
		
		\bibitem{Xia2017}
		Q.~Xia, E.~B. Sifah, K.~O. Asamoah, J.~Gao, X.~Du, and M.~Guizani, ``{MeDShare:
			Trust-Less Medical Data Sharing among Cloud Service Providers via
			Blockchain},'' \emph{IEEE Access}, vol.~5, pp. 14\,757--14\,767, 2017.
		
		\bibitem{Zhou2013}
		Z.~Zhou, H.~Zhang, X.~Du, P.~Li, and X.~Yu, ``{Prometheus: Privacy-aware data
			retrieval on hybrid cloud},'' \emph{Proceedings - IEEE INFOCOM}, no. April,
		pp. 2643--2651, 2013.
		
		\bibitem{xiao2007internet}
		Y.~Xiao, X.~Du, J.~Zhang, F.~Hu, and S.~Guizani, ``Internet protocol television
		(iptv): the killer application for the next-generation internet,'' \emph{IEEE
			Communications Magazine}, vol.~45, no.~11, pp. 126--134, 2007.
		
		\bibitem{du2008security}
		X.~Du and H.-H. Chen, ``Security in wireless sensor networks,'' \emph{IEEE
			Wireless Communications}, vol.~15, no.~4, 2008.
		
		\bibitem{zeng2018multiversion}
		Q.~Zeng, J.~Su, C.~Fu, G.~Kayas, and L.~Luo, ``A multiversion programming
		inspired approach to detecting audio adversarial examples,'' \emph{arXiv
			preprint arXiv:1812.10199}, 2018.
		
		\bibitem{yoshigoe2015overcoming}
		K.~Yoshigoe, W.~Dai, M.~Abramson, and A.~Jacobs, ``Overcoming invasion of
		privacy in smart home environment with synthetic packet injection,'' in
		\emph{TRON Symposium (TRONSHOW), 2015}.\hskip 1em plus 0.5em minus
		0.4em\relax IEEE, 2015, pp. 1--7.
		
		\bibitem{apthorpe2017smart}
		N.~Apthorpe, D.~Reisman, and N.~Feamster, ``A smart home is no castle: Privacy
		vulnerabilities of encrypted iot traffic,'' \emph{arXiv preprint
			arXiv:1705.06805}, 2017.
		
		\bibitem{apthorpe2017spying}
		N.~Apthorpe, D.~Reisman, S.~Sundaresan, A.~Narayanan, and N.~Feamster, ``Spying
		on the smart home: Privacy attacks and defenses on encrypted iot traffic,''
		\emph{arXiv preprint arXiv:1708.05044}, 2017.
		
		\bibitem{smartthingscommunity}
		\BIBentryALTinterwordspacing
		SmartThingsCommunity, ``Smartthings{P}ublic/ifttt.groovy at master ·
		{S}mart{T}hings{C}ommunity/{S}mart{T}hings{P}ublic · {G}it{H}ub,'' 2013 Feb.
		[Online]. Available:
		\url{https://github.com/SmartThingsCommunity/SmartThingsPublic/blob/master/smartapps/smartthings/ifttt.src/ifttt.groovy}
		\BIBentrySTDinterwordspacing
		
		\bibitem{webcorewiki}
		\BIBentryALTinterwordspacing
		webCoRE, ``web{C}o{RE} {W}i{K}i - {W}eb-enabled {C}ommunity's own {R}ule
		{E}ngine,'' Sep 2018. [Online]. Available: \url{https://wiki.webcore.co/}
		\BIBentrySTDinterwordspacing
		
		\bibitem{sharptools}
		\BIBentryALTinterwordspacing
		SharpTools, ``Smart{A}pp - {I}nstallation.'' [Online]. Available:
		\url{http://sharptools.boshdirect.com/installation-instructions/smartapp}
		\BIBentrySTDinterwordspacing
		
		\bibitem{casasdatasets}
		\BIBentryALTinterwordspacing
		C.~for Advanced Studies in Adaptive Systems~of WSU, ``{WSU} {CASAS}
		{D}atasets,'' 2018 March. [Online]. Available:
		\url{http://casas.wsu.edu/datasets/}
		\BIBentrySTDinterwordspacing
		
	\end{thebibliography}
	
	
	\EOD
\end{document}